\documentclass[aps,pre,preprint,superscriptaddress,longbibliography]{revtex4-2}
\usepackage{amsmath}
\usepackage{amssymb}
\usepackage{appendix}
\usepackage{bm}
\usepackage{color}

\usepackage[colorlinks=true, allcolors=blue]{hyperref}

\usepackage{graphicx}
\usepackage{subfigure}

\begin{document}

\title{Thermodynamic phase transitions in lattice spin systems with severe kinetic constraints: Numerical simulation results}

\author{Ruifeng Liu}
\email{These authors contributed equally: RL, JZ, YC}
\affiliation{
Institute of Theoretical Physics, Chinese Academy of Sciences, Zhong-Guan-Cun East Road 55, Beijing 100190, China
}
\affiliation{
  School of Physical Sciences, University of Chinese Academy of Sciences, Beijing 100049, China
}

\author{Jianwen Zhou}
\email{These authors contributed equally: RL, JZ, YC}
\affiliation{
Institute of Theoretical Physics, Chinese Academy of Sciences, Zhong-Guan-Cun East Road 55, Beijing 100190, China
}
\affiliation{
  School of Physical Sciences, University of Chinese Academy of Sciences, Beijing 100049, China
}
\affiliation{Changchun Institute of Applied Chemistry, Chinese Academy of Sciences, Changchun 130022, China}

\author{Yejia Chen}
\email{These authors contributed equally: RL, JZ, YC}
\affiliation{
Institute of Theoretical Physics, Chinese Academy of Sciences, Zhong-Guan-Cun East Road 55, Beijing 100190, China
}
\affiliation{
  School of Physical Sciences, University of Chinese Academy of Sciences, Beijing 100049, China
}

\author{Jiahang Chen}
\affiliation{
  School of Physical Sciences, University of Chinese Academy of Sciences, Beijing 100049, China
}

\author{Hai-Jun Zhou}
\email{Contact authors: YC (chenyejia@itp.ac.cn), HJZ (zhouhj@itp.ac.cn)}
\affiliation{
Institute of Theoretical Physics, Chinese Academy of Sciences, Zhong-Guan-Cun East Road 55, Beijing 100190, China
}
\affiliation{
  School of Physical Sciences, University of Chinese Academy of Sciences, Beijing 100049, China
}
\affiliation{
Institute for Advanced Physical Studies, Zhejiang University, Hangzhou 310027, China}


\date{\today}

\begin{abstract}
The Fredrickson-Andersen model with hyperparameter $K=1$ is a severely constrained kinetic lattice spin system, such that any site is temporarily blocked from changing its packing state (empty or occupied) if there is one or more occupied nearest neighbors. Starting from a completely random initial configuration with a fraction $\rho$ of sites being occupied, some of the sites may be permanently frozen to their initial state under this severe kinetic constraint. The remaining sites can switch states at least occasionally, and they form the unfrozen subsystem associated with the given initial configuration. In the present work we investigate thermodynamic phase transitions in such unfrozen subsystems of the two-dimensional square lattice and the three-dimensional cubic lattice by extensive numerical simulations. We demonstrate that the giant connected component of the unfrozen subsystem collapses at certain critical value $\rho_{c}$ of initial packing density, with $\rho_c = 0.2475$ for the square lattice and $\rho_c =  0.2809$ for the cubic lattice. This phase transition belongs to the same universality class of the conventional site percolation. We also observe that the ground states (densest packing configurations) experience a continuous crystal-to-glass phase transition at the critical value $\rho^* = 0.1423$ of initial packing density for the cubic lattice. For the two-dimensional square lattice we argue that long-range crystalline order is destroyed in the ground states as long as the initial packing density $\rho$ is positive.
\end{abstract}

\maketitle

\section{Introduction}
\label{section:introduction} 

Kinetically constrained lattice spin models have been widely adopted as simplified discrete-state systems to understand the nature of the extremely slow dynamics of glass-forming liquids and the glass transition~\cite{Ritort-Sollich-2003,Biroli-Garrahan-2013,Chacko-etal-2024}. The famous Fredrickson-Andersen (FA) model~\cite{Fredrickson-Andersen-1984} is the simplest  of such kinetic models. We may define the FA model on a finite-dimensional regular lattice, in which each site can take two possible packing states (empty or occupied), and there is a negative chemical potential $\mu$ imposed by the environment so that the individual sites prefer to be occupied. There is no any static interaction energy between two nearest-neighboring sites, but for a site to switch state it must \emph{not} be surrounded by  $K$ or more occupied nearest neighbors. This local kinetic constraint can cause very complex dynamical behaviors in the system~\cite{Fredrickson-Andersen-1984,Perrupato-Rizzo-2025b}.

Each site in a $D$-dimensional hypercubic lattice with period boundary conditions has $2 D$ nearest neighbors. If the kinetic hyperparameter $K$ is sufficiently large ($K \geq D+1$), almost all the microscopic packing configurations are mutually reachable through a kinetically allowed trajectory of single-site state flips in the thermodynamic limit of infinite system size~\cite{vanEnter-1987,Schonmann-1990,Schonmann-1992,Gregorio-etal-2004}. The thermodynamic property of the FA model is then expected to be trivial, essentially equivalent to an ideal  lattice gas of non-interacting particles.  However, if the local kinetic constraint become more severe ($K \leq D$), the microscopic configuration space will break into many different and mutually disconnected macroscopic clusters, each of which contains all the microscopic configurations that are mutually reachable through kinetically allowed single-site state flips. Each macroscopic cluster is characterized by a $K$-core of occupied sites, and these sites and some of their empty nearest neighbors are permanently frozen under the local kinetic rule. The local kinetic rule imposes a global topological constraint of $K$-core absence on the remaining unfrozen (flippable) sites, prohibiting the formation of a $K$-core of occupied sites within the subsystem of unfrozen sites~\cite{Zhou-2024}. It is now well established that this $K$-core absence topological constraint will lead to thermodynamic phase transitions in random graph ensembles, as the density of occupied sites increases beyond certain critical value~\cite{Zhou-2013,Qin-etal-2016,Guggiola-Semerjian-2015,Zhou-2022,Perrupato-Rizzo-2023}. But this thermodynamic issue has not yet been thoroughly investigated for the more realistic finite-dimensional systems.

\begin{figure}[b]
\centering
\subfigure[]{
\includegraphics[width=0.4\linewidth]{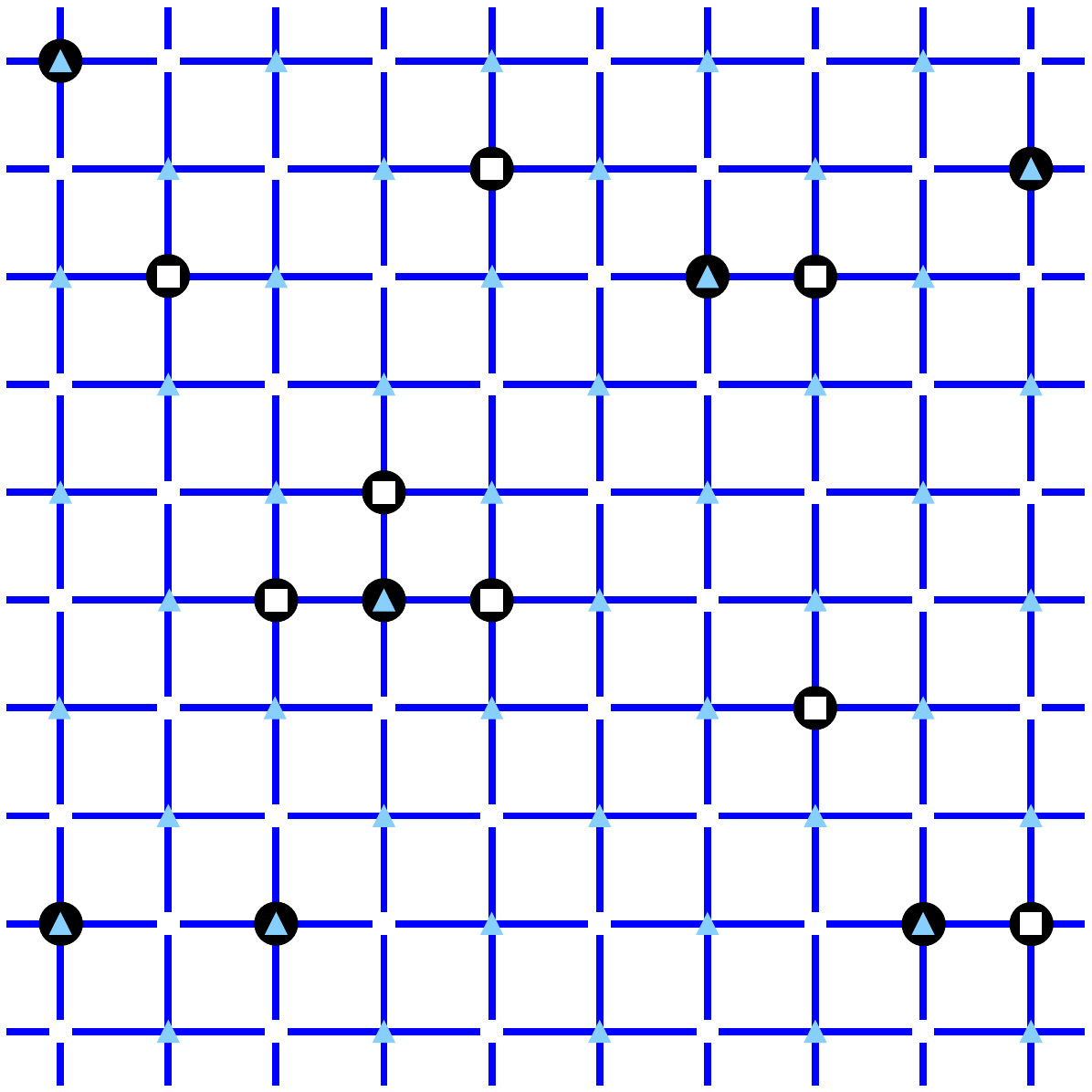}
\label{fig:FAK1model:a}
}
\hspace{0.02\linewidth}
\subfigure[]{
\includegraphics[width=0.4\linewidth]{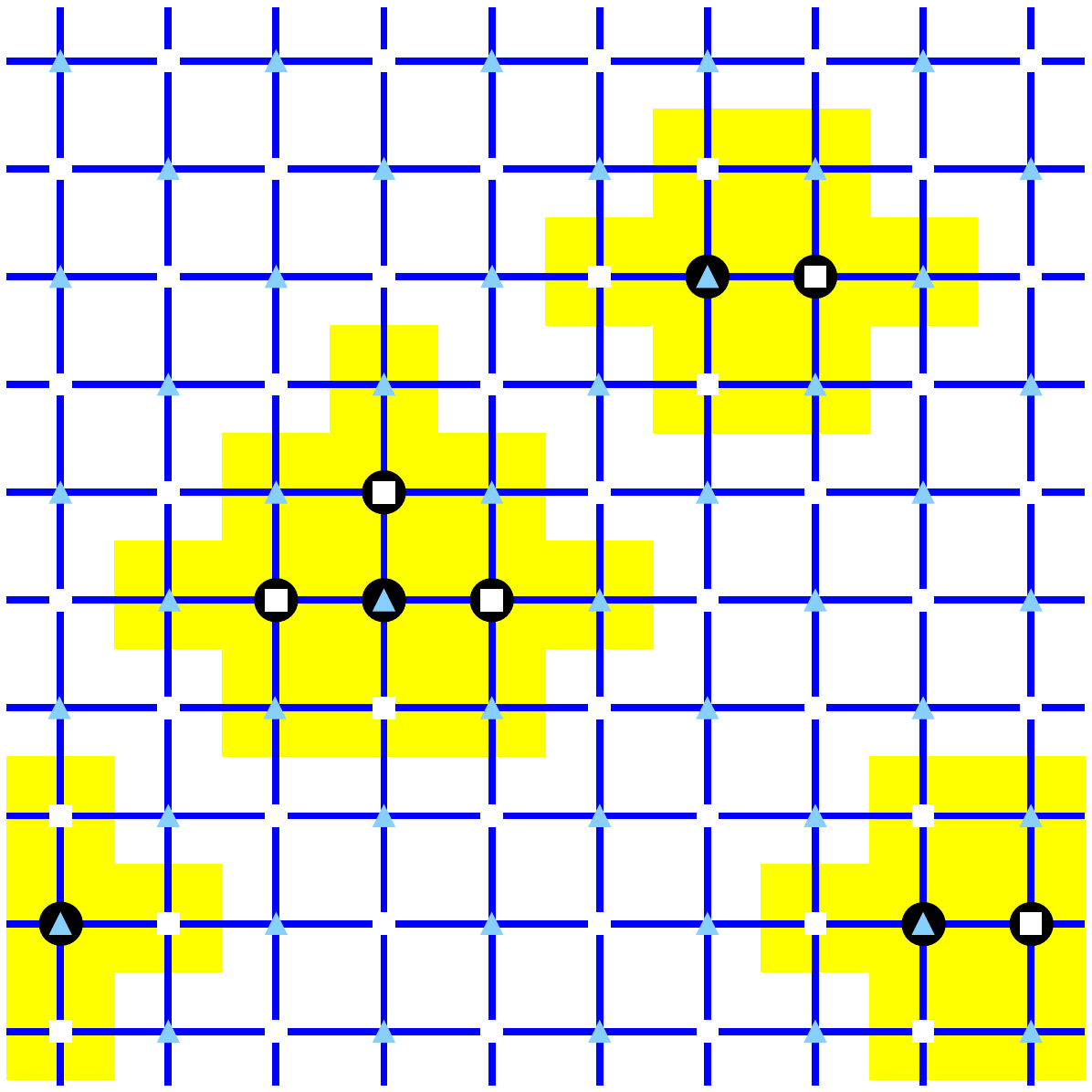}
\label{fig:FAK1model:b}
}
\caption{
The Fredrickson-Andersen model with hyperparameter $K=1$, on a square lattice of side length $L=10$ with periodic boundary condition. Sites of the two checkerboard  sublattices are distinguished by small square and triangle shapes.  (a) A random initial configuration with packing density $\rho = 0.15$. The $15$ occupied sites are indicated by large circles. (b) There are nine frozen occupied sites and  $22$ frozen empty sites, which  form three frozen domains (marked in yellow). The remaining $69$ sites belong to the unfrozen subystem and their packing states are changeable.
}
\label{fig:FAK1model}
\end{figure}

In this paper we explore possible thermodynamic phase transitions in the most severely constrained FA kinetic model with hyperparameter $K=1$, defined in Sect.~\ref{section:model}. The control parameter of our numerical simulations is $\rho$, the density (fraction) of occupied sites in the random initial packing configuration (Fig.~\ref{fig:FAK1model:a}). Because of the severe local kinetic constraint, if two occupied sites happen to be nearest neighbors, both of them will be prohibited to change state and they are then permanently frozen to the occupied initial state; according to the local kinetic rule, the empty sites that are adjacent to these frozen occupied sites are also permanently frozen to the empty initial state. These frozen occupied and frozen empty sites can be regarded as quenched defects, and we are interested in the remaining unfrozen subsystem obtained by removing these frozen sites from the original lattice (Fig.~\ref{fig:FAK1model:b}).

We first investigate the connectivity of the unfrozen subsystem (Sect.~\ref{section:percolation}). Frozen sites become more and more abundant as the density $\rho$ of initially occupied sites increases, and we find that the giant connected component of the unfrozen subsystem breaks down as $\rho$ increases to certain critical value $\rho_c = 0.2475$ for the square lattice and $\rho_c = 0.2809$ for the cubic lattice. By finite-size scaling analysis, we confirm that this continuous collapse transition belongs to the same universality class of the conventional site percolation transition~\cite{Stauffer-Aharony-1994}. This result suggests that, although the defects (frozen sites) are locally correlated in these kinetic lattice systems and they form frozen domains of various sizes and shapes, these details do not qualitatively change the nature of the percolation transition.

We then explore the effect of the frozen sites to the densest packing configurations (the ground states) of the unfrozen subsystem, within the percolating phase of $\rho < \rho_c$.  We examine square lattices and cubic lattices of various side lengths $L$. The sites in such lattice systems can be grouped into two checkerboard sublattices, such that all the nearest neighbors of a site of one sublattice belong to the other sublattice (Fig.~\ref{fig:FAK1model}). We describe in Sect.~\ref{sec:mm} how to determine the ground-state energy and how to distinguish the sites of each sublattice of the unfrozen subsystem into three categories (type-$1$, type-$0$, and type-$*$). The numerical results are then presented and discussed in   Sects.~\ref{section:phasetransition2D}, \ref{sec:ImariMa}, and \ref{section:phasetransition3D}.

For the three-dimensional cubic lattice, our numerical results offer strong evidence of a continuous crystal-to-glass phase transition at the critical initial packing density $\rho^* = 0.1429$ which is much lower than the percolation critical point $\rho_c$. For $\rho < \rho^*$ most of the occupied sites in the ground states prefer the same sublattice, so there is long-range crystalline order in a ground state; when $\rho$ exceeds $\rho^*$, however, long-range order breaks down, and the ground states of the unfrozen subsystem show properties of a glass phase: at different local regions the occupied sites prefer different sublattices, and many locally ordered domains emerge but the unfrozen subsystem as a whole loses long-range order (Sect.~\ref{section:phasetransition3D}). For the two-dimensional square lattice, our numerical results are consistent with the hypothesis of the absence of long-range order in the ground states at any positive initial packing density $\rho$; the ground-state crystal-to-glass phase transition occurs exactly at $\rho^* = 0$ (Sects.~\ref{section:phasetransition2D} and \ref{sec:ImariMa}).

Our work demonstrates that local kinetic rules can induce thermodynamic phase transitions in the finite-dimensional FA model. At $K=1$ the kinetic model is essentially the  lattice gas model with excluded-volume effect between adjacent lattice sites and with randomly distributed defects, making the system relatively easy to study. The system is also closely related with the random-field Ising model, as the frozen sites impose a local bias effect on the unfrozen subsystem. This link will be useful for further theoretical investigations~\cite{Chen-etal-2025} and for our rigorous proof (in a separate paper) that $\rho^* = 0$ in the infinite square lattice.

Another interesting extension of the present work would be to study the FA kinetic system with hyperparameter $K=2$ (slightly relaxed local kinetic constraints). The frozen occupied sites in these latter systems will form closed loops, and a microscopic configuration of the unfrozen subsystem will be a packing pattern of various tree structures of occupied unfrozen sites. The limiting initial condition of $\rho = 0$ (all sites being unfrozen) is easy to address and will soon be discussed in an accompanying paper. If the initial occupation density $\rho$ is postive, we have to consider the effects of various frozen loops. The percolation transition and the ground-state crystal-to-glass phase transition may still be preserved for the more flexible $K=2$ kinetic systems. We hope to examine the thermodynamic properties of this more challenging kinetic system in the near future. 

\section{Model}
\label{section:model}

The Fredrickson-Andersen kinetically constrained spin system~\cite{Ritort-Sollich-2003,Fredrickson-Andersen-1984,Zhou-2024,Perrupato-Rizzo-2025b} is a fundamental model for understanding the glassy dynamics of disordered systems. Given a fixed lattice structure such as a two-dimensional square lattice or a three-dimensional cubic lattice, we assign a binary packing state $c_i$ to each of its $N$ sites $i \in \{1, 2, \ldots, N\}$;  $c_i = 0$ means that site $i$ is empty and $c_i=1$ means that it is occupied (Fig.~\ref{fig:FAK1model}). A microscopic packing configuration of the whole lattice is then $\vec{\bm{c}} :=  (c_1, \ldots, c_N)$, and the total number of possible microscopic configurations is $2^N$. The total energy of each microscopic configuration is simply set to be $-\sum_i c_i$, which only depends on the total number of occupied sites. Then the completely empty configuration has the highest energy  and the completely occupied configuration has the lowest energy. Notice that this defined total energy does not contain any interaction energies. The interaction within a group of neighboring sites is implemented through the local kinetic rule. Let us denote by $b_i(t)$ the total number of occupied nearest neighbors of site $i$ at time $t$, then
\begin{equation}
    b_i(t ) \, := \, \sum\limits_{j\in \partial i} c_j(t) \; ,
    \label{eq:ni1}
\end{equation}
where $\partial i$ contains all the nearest neighboring sites of $i$ in the lattice, and $c_j(t)$ is the  state of site $j$ at time $t$. The kinetic constraint is applied to individual sites as follows: If $b_i(t) \geq K$, then site $i$ is prohibited from changing state at this time point $t$. If site $i$ has less than $K$ occupied nearest neighbors,
\begin{equation}
b_i(t) \, < \,  K \; ,
\label{eq:kineticcondition}
\end{equation}
then it is free to flip states, from $c_i = 0$ to $c_i=1$ with rate $w_{0\rightarrow 1}$ and from $c_i=1$ to $c_i=0$ with rate $w_{1\rightarrow 0}$. In this latter situation, the two flipping rates obey the detailed balance condition:
\begin{equation}
    \frac{w_{1\rightarrow 0}}{w_{0\rightarrow 1}} \, = \, e^{\mu} \; ,
    \label{eq:flipratio}
\end{equation}
where $\mu < 0$ is the chemical potential. (We can also interpret $-\mu$ as the inverse temperature and $1/(-\mu)$ as the temperature.) Conditional on site $i$ being kinetically free, then after the local flipping dynamics has reached equilibrium, the marginal probability of site $i$ being occupied will be
\begin{equation}
    p_{i} \, = \, \frac{1}{1+e^{\mu}} \; .
    \label{eq:peq}
\end{equation}
Notice that $p_{i}$ is \emph{not} necessarily the equilibrium probability of site $i$ being occupied. This is because the state flipping $c_i\rightarrow 1-c_i$ will be temporarily blocked when the flipping condition $b_i < K$ is violated~\cite{Perrupato-Rizzo-2023}.

We now introduce the initial condition, which is very important for the FA kinetic model. At time $t=0$, we assign an initial state $c_i \in \{0, 1\}$ to each site $i$, such that $c_i=1$ (occupied) with probability $\rho$ and $c_i=0$ (empty) with probability $(1-\rho)$. The initial packing states $c_i$ and $c_j$ of any two sites $i$ and $j$ are completely independent. On average the number of occupied sites at $t=0$ is $N \rho$, and we interpret the parameter $\rho$ as the density of initially occupied sites. This initial density $\rho$ is our control parameter. 

In the present work, we discuss only a single specific case of the general FA model, namely $K=1$ in the kinetic rule (\ref{eq:kineticcondition}). Sites are then temporarily blocked from switching states if they are surrounded by at least one occupied nearest neighbor (Fig.~\ref{fig:FAK1model:b}).

Given an initial random microscopic configuration with packing density $\rho$, if a site $i$ is occupied and it has one or more occupied nearest neighbors, then it will be permanently frozen to the occupied state; if a site $j$ is initially empty and it has one or more occupied nearest neighbors, its state will also be permanently frozen; if all the nearest neighboring sites of a site $k$ are empty, then site $k$ is free to flip states and it is unfrozen. Notice that the set of all the unfrozen occupied sites must form an independent set, which is a set of sites of the lattice such that all of them are not nearest neighbors; and the set of all the unfrozen empty sites is a vertex cover set, since it  contains at least one end site of every edge of the unfrozen subsystem~\cite{Hartmann-Weigt-2003}. The kinetic rule of the $K=1$ FA system therefore leads to excluded volume interactions between nearest neighboring sites of the unfrozen subsystem, making the system equivalent to the vertex cover   problem which is itself a special case of lattice glass models~\cite{Fan-Zhou-2023,Rivoire-etal-2004}. (The vertex cover problem is equivalent to the independent set problem.)

A limiting situation is $\rho = 0$, namely all the sites are empty at time $t = 0$. Then all the sites are unfrozen and they will change their states from time to time at a fixed chemical potential $\mu$, while the local kinetic rule with $K=1$ guarantees that no two nearest-neighboring sites will be occupied simultaneously. The thermodynamic property of this system is then equivalent to a hard-core lattice gas model, which has been extensively studied in the literature~\cite{Gaunt-Fisher-1965,Gaunt-1967,Bellemans-Nigam-1967}. When the underlying lattice is a two-dimensional hexagonal lattice with periodic boundary condition, such that each site has six nearest neighbors and there are three different sub-lattices, there is a continuous phase transition at a critical chemical potential $\mu_c$, between the disordered homogeneous gas phase and the symmetry-broken crystalline phase. The analytical expression of $\mu_c$ is found to $\mu_c = - \ln [ (11 + 5 \sqrt{5})/2] \approx -2.406$~\cite{Baxter-1980}. We have confirmed this prediction (data not shown) for our $K=1$ FA kinetic model under the single-flip dynamics (\ref{eq:flipratio}). In the symmetry-broken phase of $\mu < \mu_c$, the occupied sites predominantly belong to one of the three sublattices; and in the $\mu\rightarrow -\infty$ limit the ground state is  a perfect crystalline structure with maximum packing density $1/3$. 

In the present work we are interested in the range of $\rho > 0$, and we focus on the effects of the kinetically frozen sites to the remaining subsystem of unfrozen sites. 

\section{Collapse transition of the unfrozen subsystem}
\label{section:percolation}

\begin{figure*}
\centering
\subfigure[]{
\includegraphics[height=0.232\textwidth]{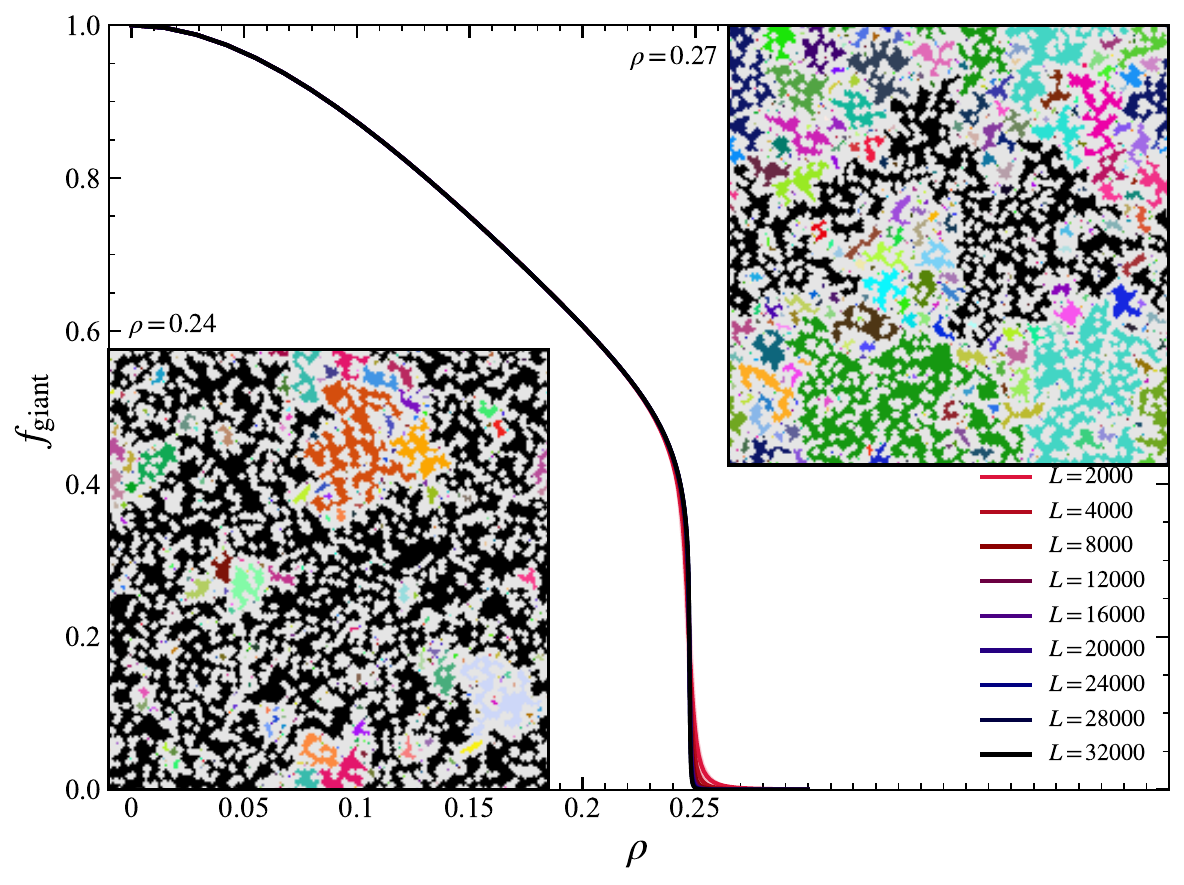}
\label{fig:rho-max_cluster-square}
}
\subfigure[]{
\includegraphics[height=0.229\textwidth]{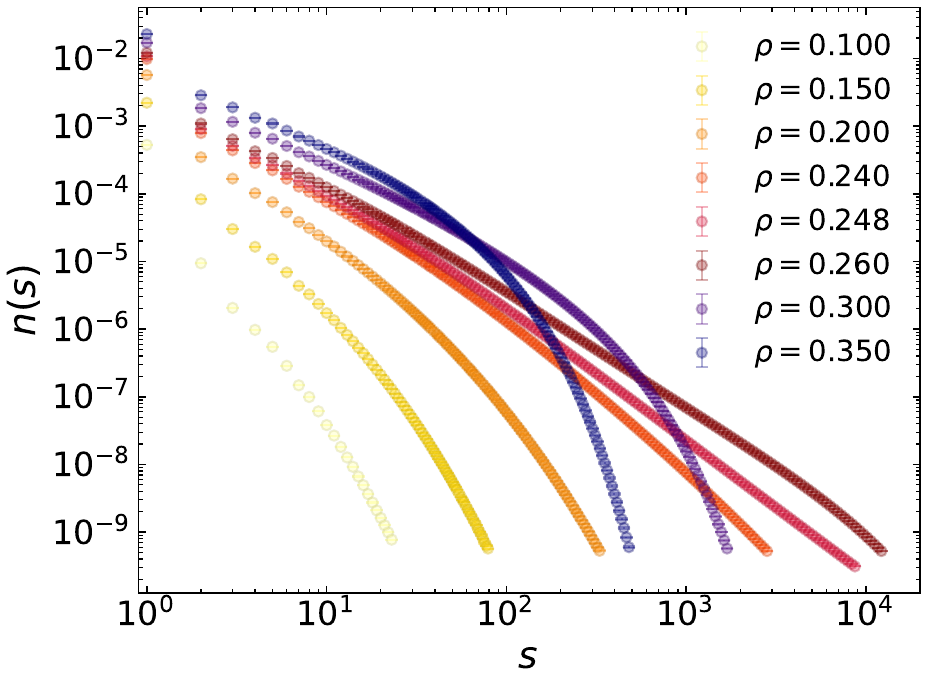} 
\label{fig:cluster-distribution-square} 
}
\subfigure[]{
    \includegraphics[height=0.231\textwidth]{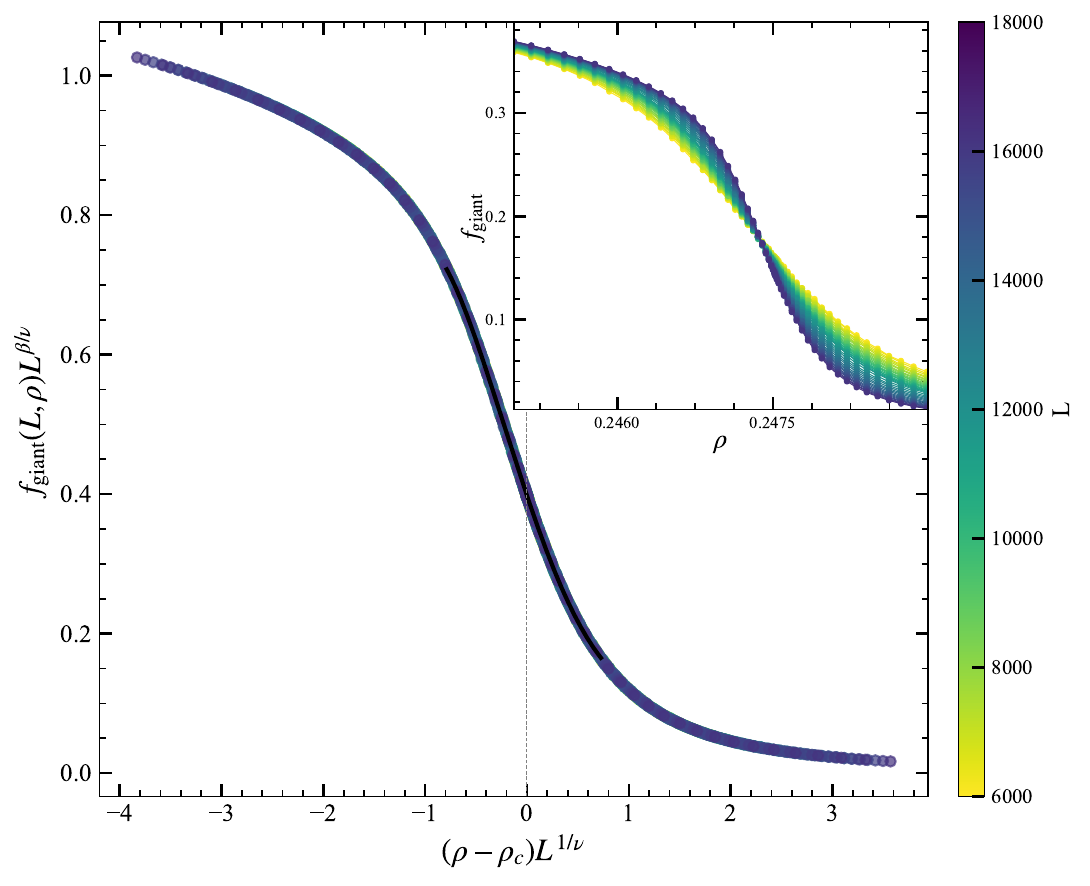}
    \label{fig:data_collapse0}
}
\caption{
Collapse transition in the periodic square lattice. (a) Relative size $f_{\textrm{giant}}$ of the giant connected component of the unfrozen subsystem versus the initial packing density $\rho$, for different side lengths $L$. Shaded areas represent standard errors of the mean. The insets show example configurations on a lattice of $L=1000$ at $\rho = 0.24$ (subcritical) and $\rho = 0.27$ (supercritical). White regions correspond to frozen sites; the other colors mark individual connected components of unfrozen sites, with the black color indicating the giant connected component.  (b) Normalized cluster size distribution $n(s)$ of unfrozen sites versus cluster size $s$ for several initial packing densities $\rho$ ($L=10000$). The distribution is roughly a power law at $\rho \approx 0.25$.
(c) Finite-size scaling analysis on $f_{\textrm{giant}}(L, \rho)$. The raw simulation data (inset) at various $L$ values collapse onto a single curve when plotted as $f_{\textrm{giant}}(L, \rho) L^{\beta/\nu}$ versus $(\rho - \rho_c)L^{1/\nu}$. The critical density is  $\rho_c = 0.247521(2)$ and the critical exponents are $\nu = 1.3375(27)$, $\beta = 0.1395(16)$.  The chi-square value of the fit is $\chi^2_{\nu} \approx 1.02$ for $1204$ degrees of freedom. 
}
\label{fig:percolation}
\end{figure*}

Starting from a random initial configuration with $N \rho$ sites being occupied, if a pair of occupied sites happen to be nearest neighbors, they will be permanently frozen to the occupied states, and in addition the nearest-neighboring empty sites of such frozen occupied sites will be frozen to the empty state. These frozen (occupied or empty) sites form many connected domains with different boundary shapes (the boundary sites of such a frozen domain must all be empty, see Fig.~\ref{fig:FAK1model:b}). All the other sites of the lattice are unfrozen and they may change state as time elapses. These unfrozen sites organize into connected components in the lattice. For example when $\rho=0$ all the sites are unfrozen, and they form a single connected component. When $\rho$ becomes considerably large, many of the sites will be frozen and the various frozen domains may break the connectivity of the unfrozen subsystem, and the  unfrozen sites will then form many small connected components.

We now investigate the collapse transition point, namely the critical initial packing density $\rho_c$ at which the giant connected component of the unfrozen subsystem breaks down~\cite{Stauffer-Aharony-1994}. To be concrete, we work on the $D=2$ square lattice and the $D=3$ cubic lattice of side length $L$ with periodic boundary condition. The total number of sites is  $N=L^D$.

Our simulation study goes as follows. (1) Generate a random packing configuration such that each site $i$ is occupied with probability $\rho$ and is  empty with probability $(1-\rho)$. (2) For each bond $(i, j)$ of the lattice, if both sites $i$ and $j$ are occupied, then denote them as frozen occupied sites; and repeat step (2) for all the bonds. (3) Then,  for each  bond $(i, j)$ of the lattice, if one of the sites (say $i$) is a frozen occupied site and the other site $j$ is empty, then denote $j$ as a frozen empty site; and repeat step (3) until all the bonds have been checked. (4) Delete all the frozen occupied and frozen empty sites from the lattice to get the unfrozen subsystem, and store the sizes of all the connected components of this unfrozen subsystem. (5) Repeat the whole process (1)-(4) starting from another random initial packing configuration. 

\subsection{Square lattices}

The simulation results on square lattices are summarized in Fig.~\ref{fig:percolation}. We perform numerical simulations for periodic square lattices up to side length $L=32000$, using $10^4$ samples (random initial configurations) for $L<1000$ and $10^3$ samples for $L \geq 1000$. We denote by $f_{\textrm{giant}}$ the fraction of sites (relative to the system size $N$) in the giant connected component of the unfrozen subsystem. As the initial packing density $\rho$ increases,  $f_{\textrm{giant}}$ decreases, and it drops dramatically at $\rho \approx 0.25$ (Fig.~\ref{fig:rho-max_cluster-square}). This behavior is characteristic of a continuous percolation transition~\cite{Stauffer-Aharony-1994}.

The size distribution $n(s)$ of connected components is of great importance in studying percolation transitions. We analyze $n(s)$ on the lattice of side length $L=10000$ (Fig.~\ref{fig:cluster-distribution-square}). At each packing density $\rho$ we compute $n(s)$ through $M = 2000$ independent numerical experiments:
\begin{equation}
n(s) \, = \,  \frac{\text{Number of components of size } s}{N \times M} \; .
\label{eq:cluster_distribution}
\end{equation}

Away from the phase transition point, $n(s)$ follows approximately a form with power-law decay and exponential cutoff:
\begin{equation}
n(s) \propto  s^{-\theta} \exp(-s/s_\xi) \; ,
\label{eq:ns_off_critical}
\end{equation}
where $\theta$ is an exponent and $s_\xi$ is a characteristic cutoff size. At the phase transition point $\rho_c$, the distribution follows a pure power law:
\begin{equation}
n(s) \propto s^{-\tau} \; ,
\label{eq:ns_critical}
\end{equation}
where $\tau$ is the Fisher exponent. Our numerical results at $\rho \approx 0.25$ yield a rough estimate of $\tau \approx 1.93$ (Fig.~\ref{fig:cluster-distribution-square}).

\subsection{Finite-size scaling analysis}

We employ finite-size scaling (FSS) analysis to determine the critical properties of the collapse transition.  Near a continuous phase  transition point $\rho_c$, the order parameter $f_{\textrm{giant}}$ is expected to follow the scaling form
\begin{equation}
f_{\textrm{giant}}(L, \rho) \, \approx \, L^{-\beta/\nu} f\bigl(  (\rho - \rho_c) L^{1/\nu} \bigr) \; ,
\label{eq:fss}
\end{equation}
where $\beta$ and $\nu$ are the critical exponents and $f(x)$ is a smooth function~\cite{Binder-1981,Hu-etal-2014,Zhu-etal-2015}. We approximate this fitting function $f(x)$ by a $q$-th order polynomial,
\begin{equation}
    f(x) = \sum_{i=0}^{q} a_i x^i \; .
    \label{eq:poly}
\end{equation}
The optimal integer value $q$ is determined by trial and error to be located in the vicinity of $q=4$. The critical parameters ($\rho_c, \beta, \nu$) and the polynomial coefficients ($a_0, a_1, \cdots, a_{q}$) are determined simultaneously by minimizing the chi-square statistic $\chi^2$, defined as
\begin{equation}
 \chi^2\, = \, \frac{
 \sum_{j=1}^Q \Bigl( \frac{ f_{\textrm{giant}}( L_j, \rho_j ) - L_j^{- \beta/\nu} f( (\rho_j - \rho_c) L_j^{1/\nu} ) }{\sigma_{(L_j, \rho_j)} } \Bigr)^2 }{Q- n_{\textrm{param}}}\; ,
 \label{eq:chisq_red}
\end{equation}
where $Q$ is the total number of data points $(L_j, \rho_j)_{j=1}^{Q}$ included in the fit, and $n_{\textrm{param}}$ is the number of free parameters (including the critical exponents and the $q+1$ polynomial coefficients); $f_{\textrm{giant}}(L_j, \rho_j)$ is taken as the empirical mean of the order parameter among all the $M$ samples at the same point $(L_j, \rho_j)$, and $\sigma_{(L_j, \rho_j)}$ is the standard error of the mean. The denominator $(Q - n_{\text{param}})$ in Eq.~(\ref{eq:chisq_red}) is the degrees of freedom of the fit.

We employ the Levenberg-Marquardt least-squares minimization algorithm to determine the optimal parameters~\cite{Levenberg-1944,Marquardt-1963}.  The quality of the fit is assessed using $\chi^2$. Data points selected for the fit are constrained to the density window $\rho \in [0.245, 0.249]$ for the $D=2$ square lattice systems. Our FSS analysis treating $\rho_c, \nu, \beta$ as free parameters yields a chi-square value $\chi^2 \approx 1.02$ (Fig.~\ref{fig:data_collapse0}), indicating an excellent fit to the scaling hypothesis. The extracted exponents are $\nu = 1.3375(27)$ and $\beta = 0.1395(16)$, with a critical density $\rho_c = 0.247521(2)$.

The critical exponents of our FSS analysis are in excellent agreement with the theoretical values for the two-dimensional percolation universality class ($\nu=4/3 \approx 1.3333$, $\beta=5/36\approx 0.1389$)~\cite{Nienhuis-etal-1979,Nijs-1979,Pearson-1980}.  We also perform another FSS analysis on our numerical data simply with fixed $\nu = 4/3$ and $\beta = 5/36$, and this second procedure produces a data collapse of virtually identical quality (with a value $\chi^2 \approx 1.04$). This stability of the fitting quality confirms that the kinetics-induced collapse transition belongs to the conventional site percolation universality class.

\subsection{Fractal dimension and the Fisher exponent}
\label{subsec:fractal dimension}

At the collapse  transition point $\rho = \rho_c$, we see from Eq.~(\ref{eq:fss}) that the relative size of the giant connected component of unfrozen sites is $f_{\textrm{giant}} \propto  L^{-\beta/\nu}$, with $\beta/\nu \approx 0.1042$. This indicates that at the critical point $\rho_c$, the number of sites in the giant connected component, $S_{\textrm{giant}} = N f_{\textrm{giant}}$, will follow the scaling relationship
\begin{equation}
S_{\textrm{giant}}(L) \, \propto \, L^{D_{\textrm{giant}}} \; , 
\label{eq:fss_mmax}
\end{equation}
with a fractal dimension $D_{\textrm{giant}} = 2 - \frac{\beta}{\nu} \approx 1.8958$. The Fisher exponent $\tau$ of Eq.~(\ref{eq:ns_critical}) for the component size distribution $n(s)$ at the critical point is computed to be $\tau \approx 2.0549$ by using the following relationship
\begin{equation}
\tau \, = \, \frac{2}{D_{\textrm{giant}}} + 1 \; .
\label{eq:hyperscaling}
\end{equation}
This expression (\ref{eq:hyperscaling}) can be understood as follows. By the definition (\ref{eq:cluster_distribution}), which has the rescaling factor $N$, we see that the mean value of the giant connected component size $S_{\textrm{giant}}$ satisfies
\begin{equation}
    \int_{S_{\textrm{giant}}}^{\infty} n(s) \textrm{d} s \, = \, \frac{1}{N} \; ,
\end{equation}
because the number of connected components with size $\geq S_{\textrm{giant}}$ should be unity in the average sense. At the critical point $\rho_c$, $n(s)$ is following the power-law (\ref{eq:ns_critical}), and therefore we obtain from the above expression that $N S_{\textrm{giant}}^{1-\tau}  = \textrm{constant}$. In the limit of $N\gg 1$, we see that $\tau - 1 = \ln N / \ln S_{\textrm{giant}} = 2/D_{\textrm{giant}}$, which is Eq.~(\ref{eq:hyperscaling}). 

We have confirmed Eq.~(\ref{eq:fss_mmax}) following the same finite-size scaling method of the preceding subsection, obtaining $D_{\textrm{giant}} = 1.8956(3)$ for $D=2$ which is in excellent agreement with the theoretical value $91/48\approx 1.8958$ of two-dimensional site percolation~\cite{Harrison-etal-1978,Stauffer-1979}. The exponent $D_{\textrm{giant}}$, which is slightly below two, reveals the fractal property of the giant connected component of the unfrozen subsystem. 

\subsection{Cubic lattices}

\begin{figure*}
\centering
\subfigure[]{
\includegraphics[height=0.235\textwidth]{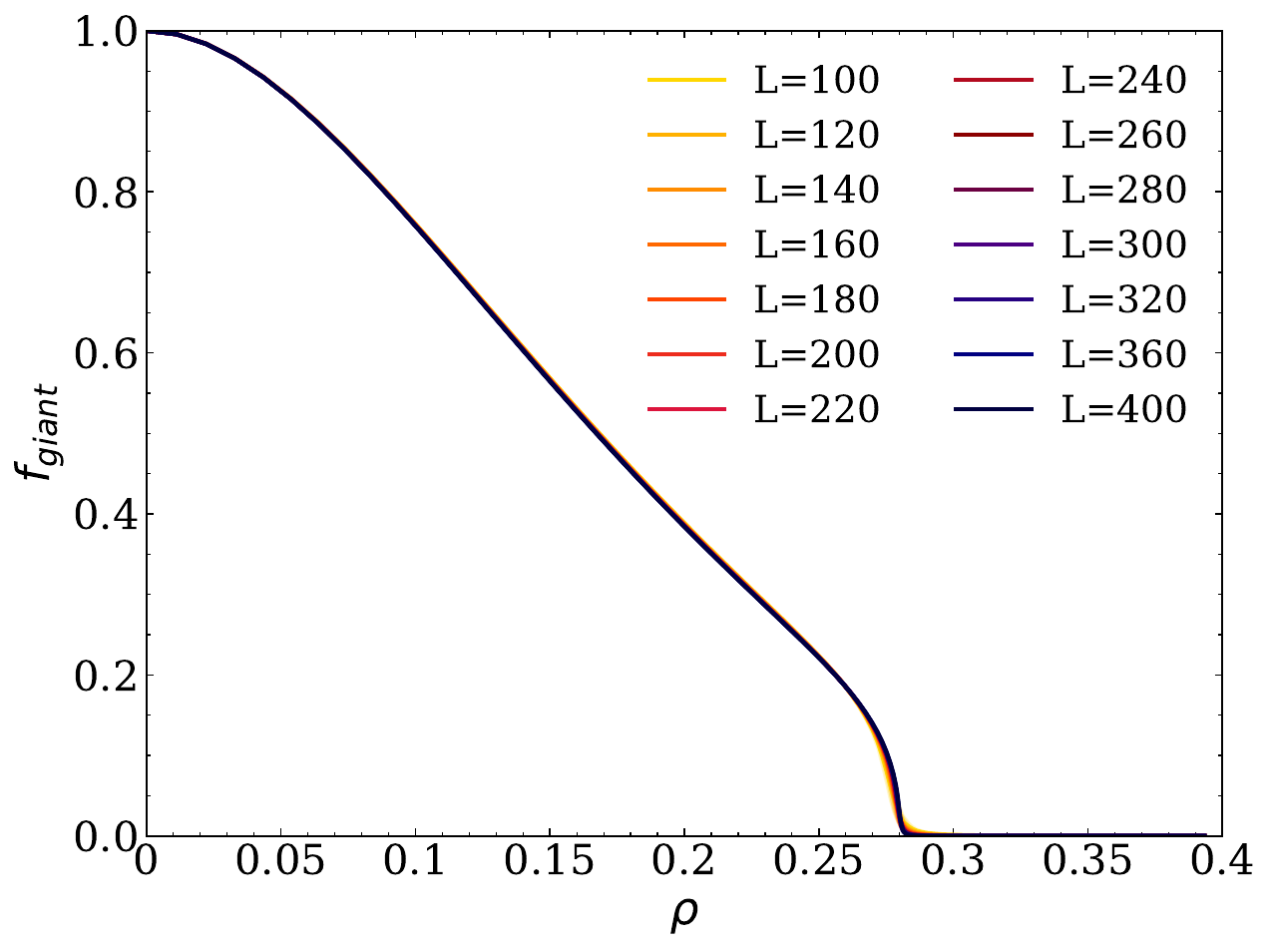}
\label{rho-max_cluster-cubic}
}
\subfigure[]{
\includegraphics[height=0.235\textwidth]{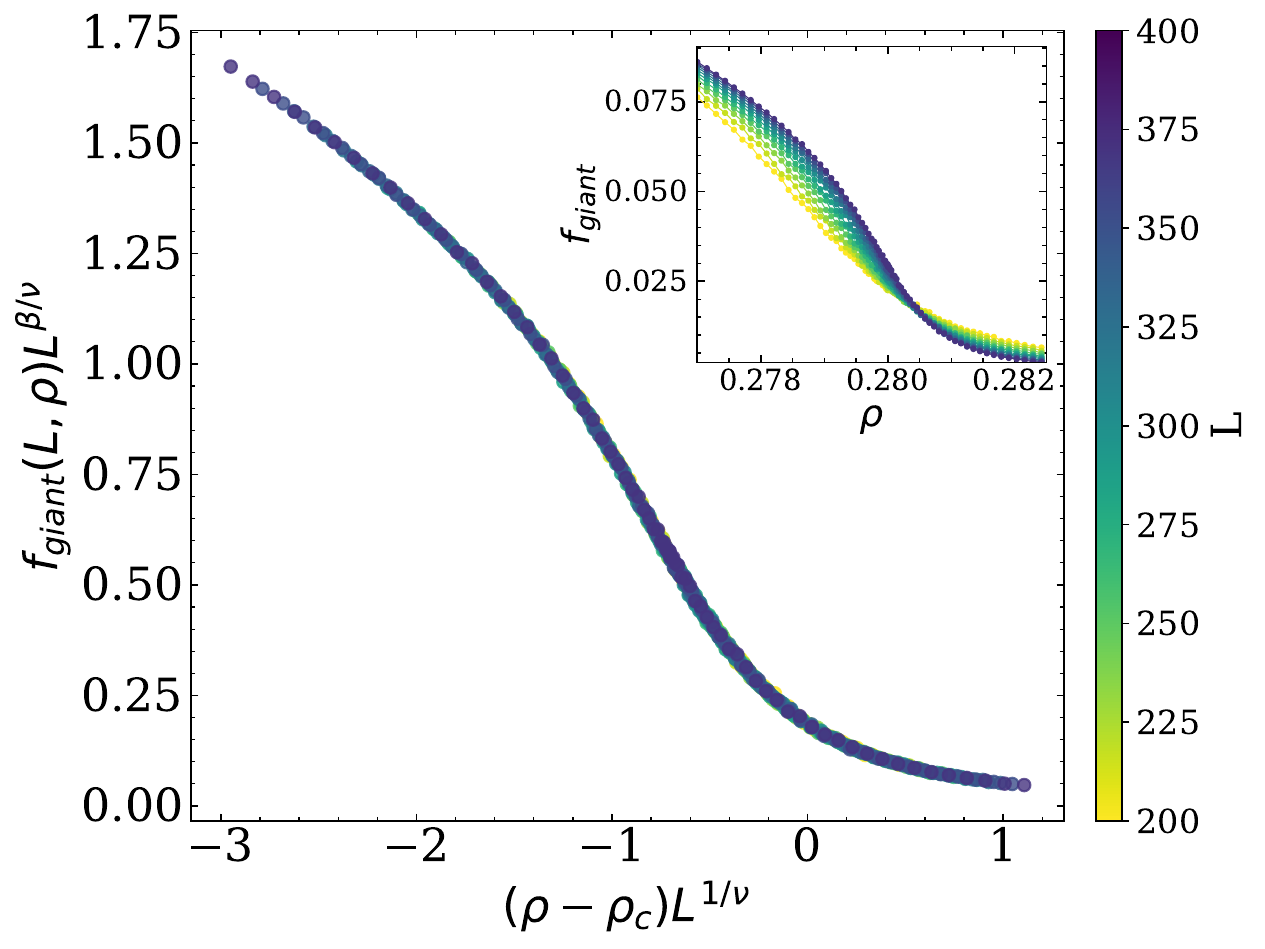}
\label{fig:data_collapse_3D_percolation}
}  
\caption{
Same as Figs.~\ref{fig:rho-max_cluster-square} and \ref{fig:data_collapse0}, but for the periodic cubic lattices ($D=3$) of various side lengths $L$. (a) The giant connected component of the unfrozen subsystem collapses at $\rho \approx 0.28$. (b) Finite-size scaling analysis of $f_{\textrm{giant}}(L, \rho)$. The critical density is $\rho_c = 0.280942(4)$ and the critical exponents are $\nu = 0.9053(33)$, $\beta = 0.4482(19)$.  The chi-square value of the fit is $\chi^2_{\nu} \approx 0.97$ for $841$ degrees of freedom.
}
\label{fig:percolation_cubic}
\end{figure*}

We also perform percolation analysis on the three-dimensional periodic cubic lattice. Lattice side lengths up to $L=400$ are considered, and the number of independent samples (random initial configurations) is set to be $10^4$ for $L < 100$ and $10^3$ for $L \geq 100$. 

The giant connected component of the unfrozen subsystem collapses at $\rho \approx 0.28$ (Fig.~\ref{rho-max_cluster-cubic}), and this collapse transition is also continuous as in the two-dimensional square lattice. A finite size scaling analysis is performed to determine precisely the critical point of this transition (Fig.~\ref{fig:data_collapse_3D_percolation}). Data points selected for the fit are constrained to the density window $\rho \in [0.277, 0.283]$. Our fitting yields a chi-square value $\chi^2\approx0.97$, indicating a very good fit to the scaling hypothesis (\ref{eq:fss}). The extracted critical exponents are $\nu = 0.9053(33)$ and $\beta = 0.4482(19)$, and the critical density is $\rho_c = 0.280942(4)$.

The quantitative results of our FSS analysis are quite close to the modern estimation for the three-dimensional site percolation universality class ($\nu=0.88(2)$, $\beta=0.429(4)$~\cite{surmonte1976,PhysRevB.41.9183,xusimultaneous2014,PhysRevD.103.116024,GIMENEZ2025130562}). Using our estimated values of $\beta$ and $\nu$, we obtain from Eq.~(\ref{eq:hyperscaling}) an estimate Fisher exponent value  $\tau= 2.1976(13)$, which is in good agreement with the modern estimation of $\tau=2.1938(12)$~\cite{xusimultaneous2014,PhysRevD.103.116024}. These comparative results suggest that the three-dimensional collapse phase transition again belongs to the same universality class of the site percolation transition.

\section{The ground-state subspace}
\label{sec:mm}

The preceding section confirms the existence of a continuous collapse phase transition induced by the kinetically frozen sites. We continue to explore the effects of the frozen sites to the ground states (the densest packing configurations) of the unfrozen subsystem. The density $\rho$ of occupied sites in the random initial configurations is chosen to be less than the percolation threshold value $\rho_c$, so that the unfrozen subsystem has an extensive giant connected component. The ground states correspond to the chemical potential being $\mu=-\infty$ in Eq.~(\ref{eq:flipratio}).  In the next two sections we will examine periodic square lattices ($D=2$) and cubic lattices ($D=3$). Here we describe our numerical strategy.

\subsection{The ground-state energy}

Given an unfrozen subsystem, we define its energy as
\begin{equation}
E \, = \, -\sum_j{^\prime} c_j \; ,
\label{Eq:energy_function_2}
\end{equation}
where the symbol $\sum^\prime$ means that the summation is over all the unfrozen sites $j$.  If an unfrozen site $j$ changes from being empty ($c_j=0$) to being occupied ($c_j=1$), the total energy $E$ decreases by one unit, so the occupied state is energetically favored. A  ground state is then a lowest-energy densest packing configuration of all the $N_{\textrm{uf}}$ unfrozen sites, such that no two nearest neighboring sites are occupied simultaneously. We denote by $N_{\textrm{uf}}$ the total number of unfrozen sites and by $E_{\textrm{min}}$ the ground-state energy ($- N_{\textrm{uf}} < E_{\textrm{min}} < 0$).

The occupied sites in any ground state of the energy function (\ref{Eq:energy_function_2})  form a maximum independent set and the empty (unoccupied) sites form a minimum vertex cover set. Given a generic graph formed by a set of sites and a set of edges between these sites, an   independent set is a subset of sites such that there is no edge between any two sites of this subset, and a vertex cover is a subset of sites which contains at least one end site of any edge of the graph. Exactly  solving the maximum independent set problem (and the equivalent minimum vertex cover problem) is in general an extremely difficult combinatorial optimization problem~\cite{Hartmann-Weigt-2003,Zhou-2005a}, but for the $D$-dimensional hypercubic lattices, single maximum independent set solutions can be constructed by solving the easy maximum matching problem, which aims at creating an edge set containing the maximum number of mutually separated edges (no two edges of this set share the same end site) ~\cite{Ogielski-1986,Hartmann-2011}. This equivalence follows from the  bipartite structural property of hypercubic lattices~\cite{Rizzi-2000}.

\begin{figure}
  \centering
  \includegraphics[width=0.6\linewidth]{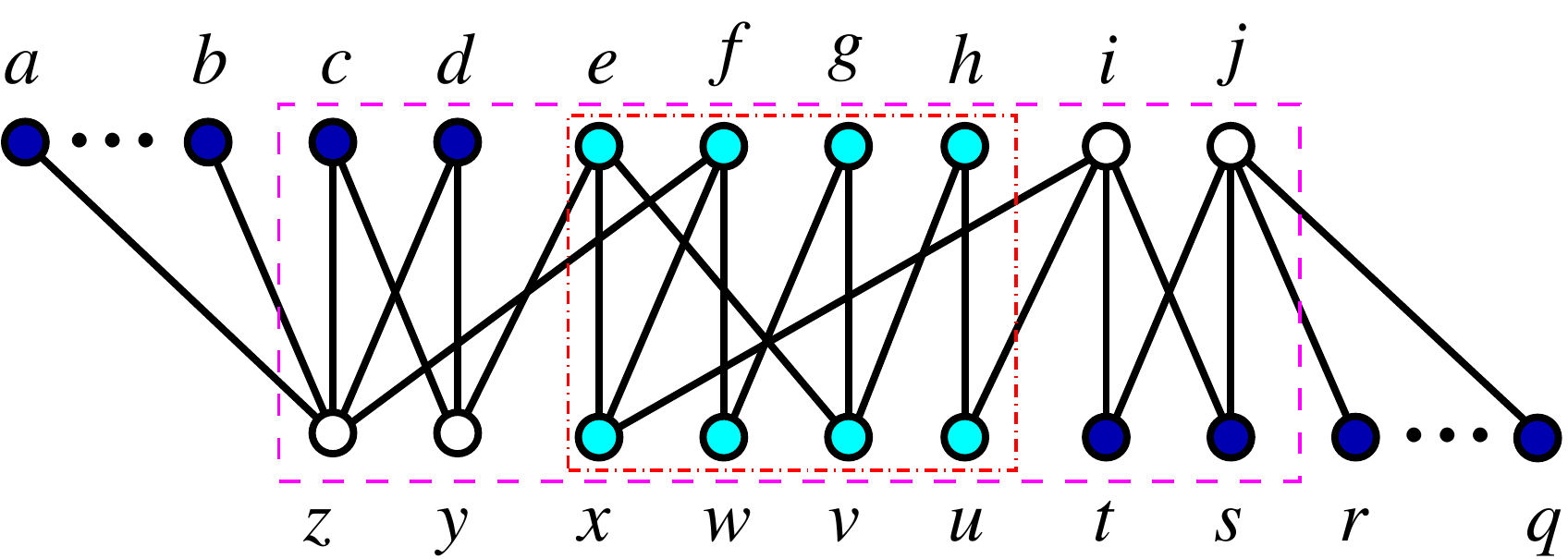}
  \caption{
  An example small  bipartite graph. The top and bottom rows correspond to sublattices A and B, respectively. The eight vertical edges $(c, z), \ldots, (j,s)$
  form a maximum matching solution $\textrm{MM}$. The subgraph induced by all the sites of $\textrm{MM}$ is marked by the dashed rectangle. The state-flexible subgraph  as obtained through a fix-and-propagate process is indicated by the dot-dashed rectangle. The type-$1$, type-$0$, and type-$*$ sites are distinguished by blue, white, and cyan circles, respectively.
}
\label{f120226}
\end{figure}

Given a random initial packing configuration with packing density $\rho$, we first specify all the unfrozen sites of the lattice and  represent the whole unfrozen subsystem as a bipartite graph $G$. Some of the sites of graph $G$ belong to sublattice A of the hypercubic lattice and the other sites  belong to sublattice B, and all its edges are linking between these two sublattices. We then apply a very efficient local search dynamic (such as the Hopcroft-Karp algorithm) on this bipartite system $G$ to obtain a single maximum-matching solution~\cite{Hopcroft-Karp-1973,Zhou-Ouyang-2003}. Let us denote this solution as $\textrm{MM}$. For ample, the edge set $\textrm{MM} = \{\ (c, z),\ (d, y),\ \ldots,\ (i, t),\ (j, s)\ \}$ is a maximum matching of the small bipartite graph in Fig.~\ref{f120226}.

We refer to all the edges of the maximum matching  $\textrm{MM}$ as the matching edges and all the other edges (such as edges $(d, z)$ and $(j, q)$ of Fig.~\ref{f120226}) as the unmatched edges. All the end sites of these matching edges are referred to as matched sites and all the other sites are unmatched sites. Notice that if two sites both are unmatched, then there is no edge between them. These two sites can belong to the same sublattice (such as sites $a$ and $b$ of Fig.~\ref{f120226}) or belong to the two different sublattices (such as sites $a$ and $r$). 

After obtaining the maximum-matching $\textrm{MM}$, we can then determine the ground-state energy of Eq.~(\ref{Eq:energy_function_2}) as
\begin{equation}
E_{\textrm{min}} \, = \, - N_{\textrm{uf}} +  \bigl| \textrm{MM} \bigr| \; ,
\label{eq:Emin}
\end{equation}
where $| \textrm{MM} |$ is the the maximum-matching size. This expression is known as the K\"{o}nig theorem in the literature~\cite{Rizzi-2000}. We will present a simple constructive proof in the next subsection.

\subsection{Three types of unfrozen sites}
\label{subsec:3type}

To have a global and comprehensive view of the whole configuration subspace formed by all the ground states of the unfrozen subsystem, we will classify all the unfrozen sites into three types: type-$1$ (being occupied in all the ground states); type-$0$ (being empty in all the ground states); and type-$*$ (being occupied in some ground states and empty in the other ground states)~\cite{Zhou-2005a,Wei-etal-2015,Li-etal-2026}. We carry out this classification task, together with the construction of a densest packing configuration,  under the guidance of the single maximum matching solution $\textrm{MM}$ for the bipartite graph $G$.

We now explain the four major steps using the small bipartite graph of Fig.~\ref{f120226} as a concrete example. 

\begin{enumerate}
\item[(1)] We set all the unmatched sites ($a, \ldots, b$ and $r, \ldots, q$) to be occupied and put them to the set of type-$1$ sites; then, we set all the nearest neighbors of these fixed  sites to be empty and put them to the set of type-$0$ sites (all of them must be matched sites of $\textrm{MM}$, such as $z$ and $j$).
\item[(2)] We set the matched nearest neighboring sites (such as $c$ and $s$) of these empty matched sites to be occupied and add them to the set of type-$1$ sites; and then for these newly fixed occupied sites, we set all their other nearest neighbors (whose states are not yet fixed, such as sites $y$ and $i$) to be empty and add them to the set of type-$0$ sites.
\item[(3)] We then keep repeating step (2), until no other sites can be fixed by this propagation process.
\end{enumerate}
Notice that the propagation processes from left (starting from sites $a, \ldots, b$) and from right (starting from sites $q, \ldots, r$) will never cause a conflict, with the left propagation requiring a site (say $d$) of sublattice A to be occupied while the right propagation requiring this same site $d$ to be empty. Because if such a conflict occurs, it means that there is a path of odd length $2 k + 1$ (with $k \geq 1$) linking an unmatched site (say $a$) of sublattice A with an unmatched site (say $q$) of sublattice B, with all the $2 k - 1$ internal edges of this path belonging to set $\textrm{MM}$; we can then increase the maximum matching size by one, simply by deleting all the matching edges of this path from set $\textrm{MM}$ and then add all the unmatched edges of this path to $\textrm{MM}$; but this is in contradiction with the premise that the original edge set $\textrm{MM}$ has already been a maximum matching.

We will refer to the bipartite subgraph of unfixed sites after this whole fix-and-propagate process as the \emph{state-flexible subgraph}. This subgraph of $G$ contains some matching edges, and it may also contain some unmatched edges. For the example in Fig.~\ref{f120226} the state-flexible subgraph is the one induced by all the sites in the set $\{e, f, g, h, u, v, w, x\}$.

\begin{enumerate}
\item[(4)] There are at least two ways of setting the packing states of all the sites in the state-flexible subgraph. For example, we can set all the sites of sublattice A in this subgraph ($e, f, g, h$ of Fig.~\ref{f120226}) to be occupied and all the sites belonging to sublattice B ($u, v, w, x$) to be empty, or the complete opposite. We put all the sites of the state-flexible subgraph to the set of type-$*$ sites.
\end{enumerate}
It is easy to check that the resulting whole set of occupied sites must be a maximum independent set of the bipartite graph $G$. The total number of sites in this set is $N_{\textrm{uf}} - |\textrm{MM}|$, so we have Eq.~(\ref{eq:Emin}).

For any matched edge $(g, v)$ in the given  maximum matching solution $\textrm{MM}$, the end sites $g$ and $v$ are mutually exclusive and predictive in any single maximum independent set solution: if site $g$ is occupied then site $v$ must be empty, and if site $g$ is empty then site $v$ must be occupied~\cite{Wei-etal-2015}. To prove this important   one-and-only-one property, we may notice that the complement of a maximum independent set is a minimum vertex cover set (say $\textrm{MVC}$), which contains all the empty sites as fixed during the above-mentioned construction steps (1)-(4). For any edge of the bipartite graph $G$, at least one of the two end sites must belong to this $\textrm{MVC}$. Because the size of $\textrm{MVC}$ is equal to that of $\textrm{MM}$, then one and only one of the two end sites $g$ and $v$ of any matched edge $(g,v)$ of $G$ will be found in $\textrm{MVC}$, which also means that one and only one of them will be found in any maximum independent set.

To prove that all the type-$1$ sites as determined during the above-mentioned steps (1)-(3) must be occupied in all the ground states of the unfrozen subsystem, we first notice that all the unmatched sites with respect to $\textrm{MM}$ must belong to all the maximum independent set solutions~\cite{Wei-etal-2015}. If an unmatched site (say $a$ of Fig.~\ref{f120226}) is not a member of an independent set, this independent set can not be a maximum independent set, because the complementary vertex cover set to this independent set can not be a minimum vertex cover (it must contain this unmatched site $a$ in addition to $|\textrm{MM}|$ matched sites). With all the unmatched sites being justified as type-$1$ sites, then all the nearest neighbors of these sites must belong to each and every minimum vertex cover solution of the bipartite graph $G$,  and therefore they are justified as type-$0$ sites. Following this same  reasoning, all the sites fixed during the above-mentioned iterative fix-and-propagate process (1)-(3) can be justified as type-$1$ or type-$0$ sites of the unfrozen subsystem. All the sites of the remaining state-flexible subgraph of $G$ are obviously justified as type-$*$ sites.

\subsection{Uniform sampling of ground states}
\label{subsec:usgs}

The type-$*$ sites of the state-flexible subgraph contribute to the entropy of the ground states of the unfrozen subsystem.  The ground-state entropy was found to be extensive for random graphs~\cite{Zhou-Zhou-2009,Wei-etal-2015,Wei-Zhang-etal-2015}. We expect that the ground-state entropies should  also be extensive for the present lattice systems, since the number of type-$*$ sites is proportional to the total number of unfrozen sites (see below).  About half of these type-$*$ sites may be in the occupied state in a randomly sampled ground-state configuration. But individual type-$*$ sites may have different probabilities of being occupied among all the ground-state configurations. There may exist extremely strong correlations between the states of two type-$*$ sites which are not connected by a matching edge (e.g., $e$ and $v$ of Fig.~\ref{f120226}), such that the state of one site can completely predict the state of the other site~\cite{Zhou-2005a}. We can take advantage of these extreme correlations to simplify the state-flexible subgraph by a cycle-simplification process (see Appendix~\ref{app:cs} for technical details)~\cite{Wei-etal-2015,Wei-Zhang-etal-2015}.

We have performed conventional Markov-Chain Monte Carlo simulations to sample the ground-state configuration space of the simplified state-flexible subgraph uniformly at random~\cite{Wei-Zhang-etal-2015}. After obtaining a large set of individual ground states, we have then examined the degrees of dominance of one sublattice A over the other sublattice B. We have found that the numerical results are almost indistinguishable from the numerical results obtained employing the  ``coarse-grained" type-$1$, type-$0$,  and type-$*$ states.

A huge advantage of these coarse-grained states is that we need only to access a single maximum matching solution for each sample of initial packing configurations. In the next two sections \ref{section:phasetransition2D} and \ref{section:phasetransition3D}, therefore, we only present the numerical results obtained on the coarse-grained states.

\section{Ground states of square lattices}
\label{section:phasetransition2D}

\begin{figure}
\centering
\subfigure[]{
\includegraphics[width=0.5\linewidth]{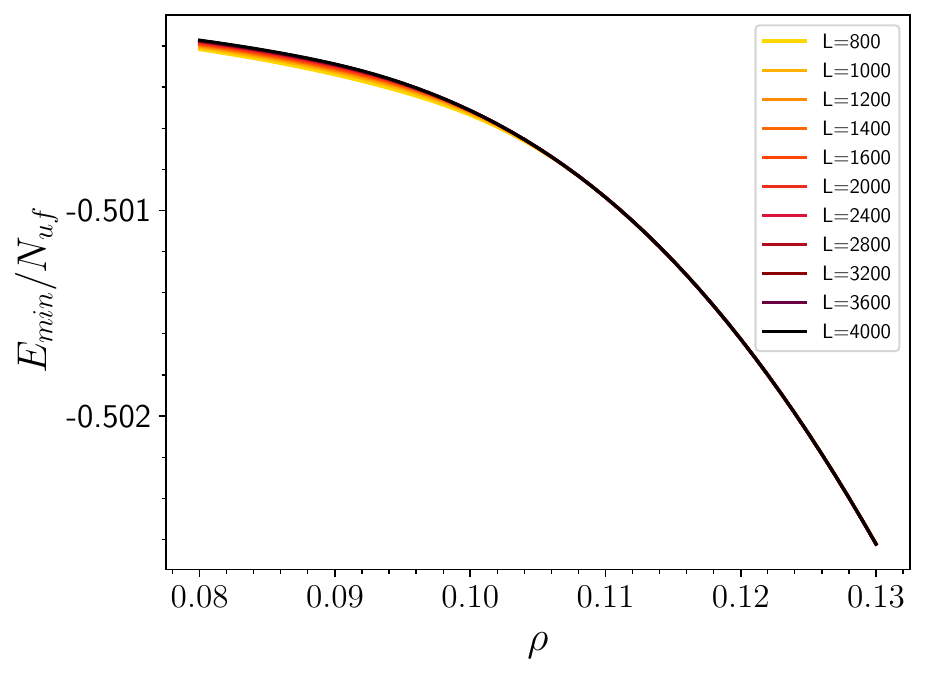}
\label{fig-E_mean_vs_rho_d2}
}

\subfigure[]{
\includegraphics[width=0.25\linewidth]{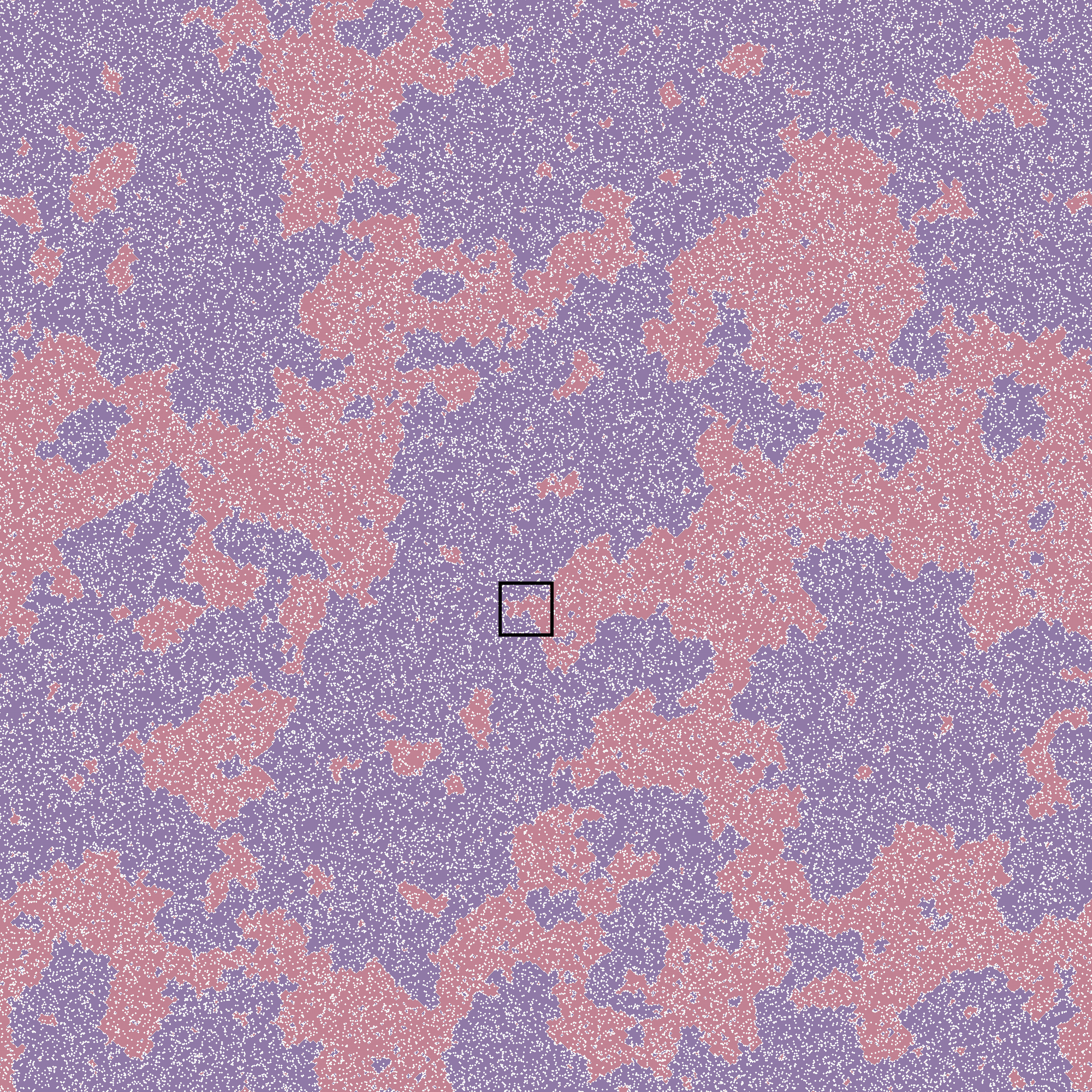}
\label{fig-GS-2d-square}
}
\hspace{0.1\linewidth}
\subfigure[]{
\includegraphics[width=0.25\linewidth]{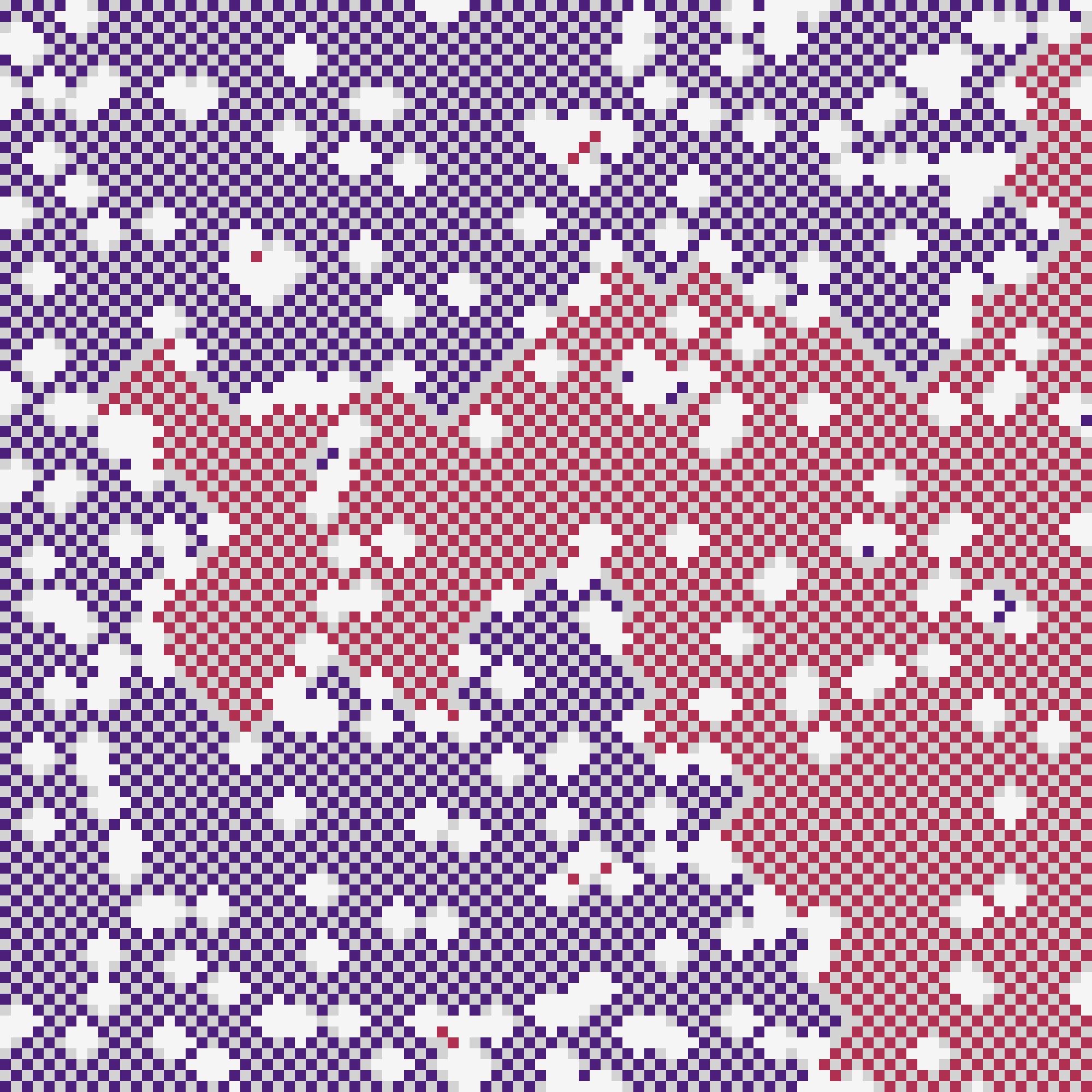}
\label{fig-GS-2d-square-3}
}
\caption{
(a) The ground-state energy density $E_{\textrm{min}}/N_{\textrm{uf}}$ versus the initial density $\rho$ of occupied sites, for the periodic square lattices of various side lengths $L$. Each data point is the average over $1000$ independent samples and shaded regions represent standard errors of the mean. (b) A single ground state example for a system of side length $L=2000$ at $\rho = 0.12$, with energy density $E_{\textrm{min}}/N_{\textrm{uf}} = -0.5016$. Occupied unfrozen sites of the two sublattices are distinguished by blue and red colors, empty unfrozen sites are marked by gray color, and the frozen sites are marked by white color. Panel (c) is highlight of the boxed $100\times 100$ region of (b).
}
\end{figure}
\begin{figure*}
\centering
\subfigure[]{
\includegraphics[width=0.45\textwidth]{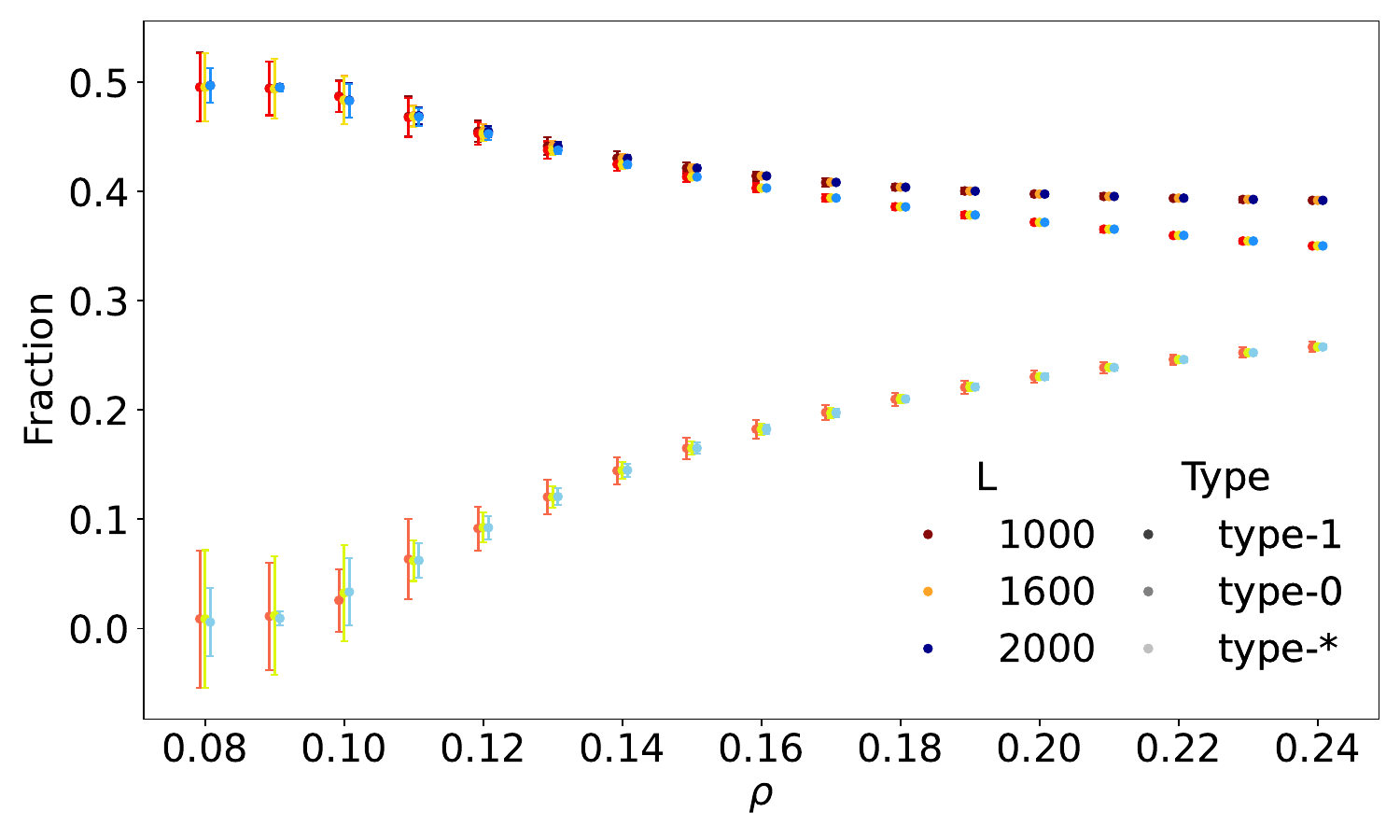}
\label{fig-GS-space_2D_joint}
}
\subfigure[]{
\includegraphics[width=0.45\textwidth]{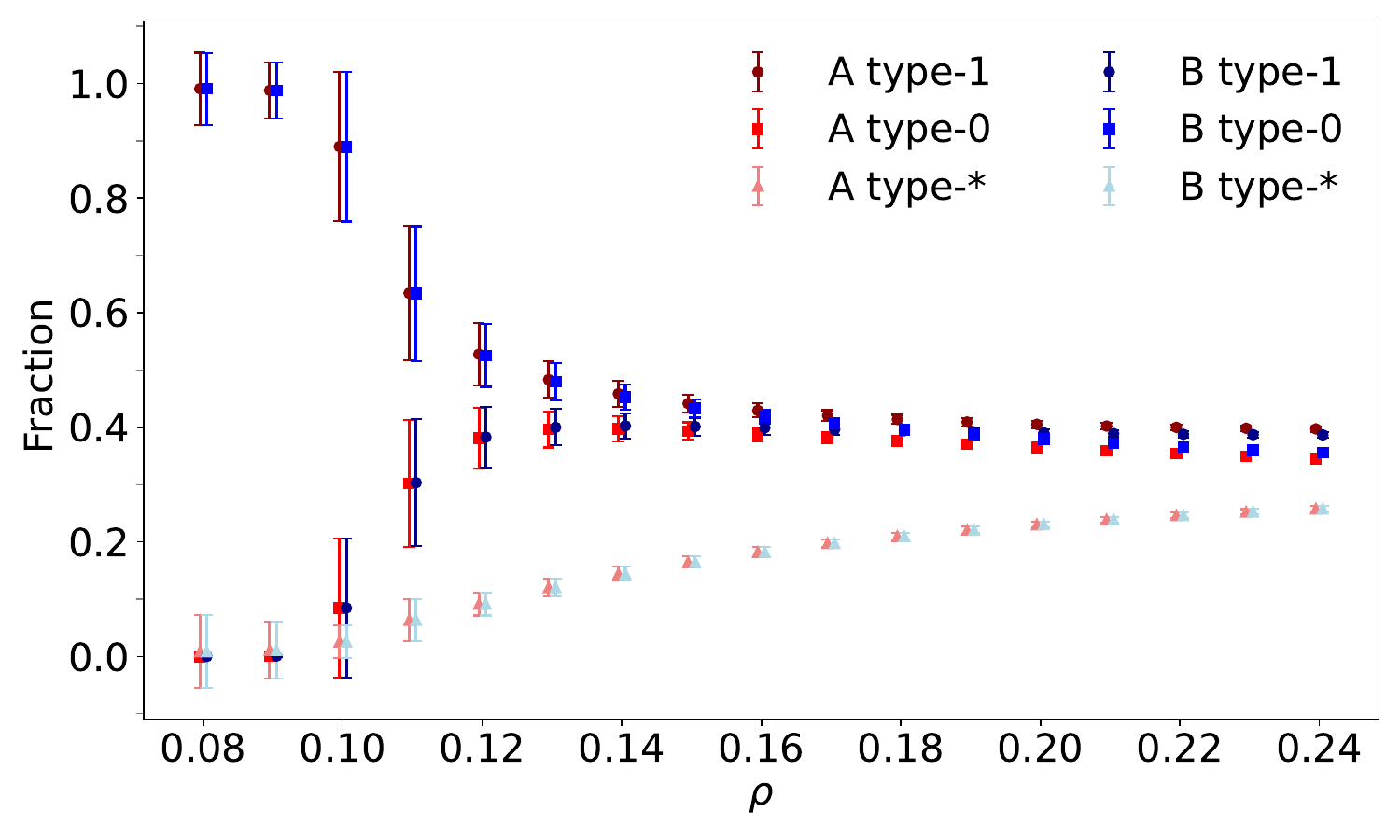}
\label{fig-GS-space_2D_AB}
}
\caption{
Abundance of type-$1$, type-$0$, and type-$*$ sites in the unfrozen subsystem of the periodic square lattices. (a) Fractions of the three types of sites within the giant connected component of the unfrozen subsystem. Lattice side lengths $L=1000$, $1600$, and $2000$. (b) Fractions of the three types of sites within a single sublattice A or B of the giant connected component of the unfrozen subsystem ($L=1000$). For each initial packing density $\rho$, the mean and standard deviation are evaluated from $1000$ independent samples of initial packing configurations. A small horizontal offset ($0.0005$) is applied to the data points at the same $\rho$ values for clarity.
}
\label{fig-GS_space_2D}
\end{figure*}

We first apply the method of Sec.~\ref{sec:mm} to the two-dimensional periodic square lattices, in which every site has four nearest neighbors. At each point $(L, \rho)$ of lattice side length $L$ and initial packing density $\rho$, we sample $1000$ independent random initial packing configurations and obtain the corresponding $1000$ unfrozen bipartite graphs $G$ of slightly different sizes $N_{\textrm{uf}}$. A single maximum matching solution $\textrm{MM}$ is then constructed for each of these bipartite graphs, and the ground-state energy $E_{\textrm{min}}$ is computed through Eq.~(\ref{eq:Emin}). 

Figure~\ref{fig-E_mean_vs_rho_d2} shows the mean ground-state energy densities $E_{\textrm{min}}/N_{\textrm{uf}}$ for various system sizes $L$. If all the sites are initially unoccupied ($\rho=0$),  there is no frozen disorder and so the ground-state energy density is exactly $-1/2$, with one sublattice A being completely occupied and the other sublattice B being completely empty. As the initial packing density increases, we find that the ground-state energy density deviates more and more from $-1/2$, but it appears to be a smooth function of $\rho$ (see Fig.~\ref{fig-E_mean_vs_rho_d2}, which highlights the range of $0.08 \leq \rho \leq 0.13$).

\subsection{Competition between the two sublattices}
\label{subsec:competition}

We present in Fig.~\ref{fig-GS-2d-square} one concrete example of ground states for a square lattice of side length $L=2000$,  obtained at $\rho = 0.12$. This ground state clearly displays an amorphous structure. In some regions one of the two checkerboard sublattices ($A$) is almost completely occupied while the other sublattice ($B$) is almost completely unoccupied; in some other regions the packing pattern is the opposite. We have checked that as $\rho$ increases, the individual regions of occupied sublattice will shrink in size, and the interfaces between these occupied unfrozen sublattices becomes more and more complex.

The breakdown of the global dominance of one single sublattice can be understood as follows. Frozen sites act as defects and they effectively impose local biases to the adjacent unfrozen sites. Within each local region these frozen sites usually favor one of the two checkerboard sublattices to be occupied, even at the expense of forming an interface (domain wall) with a neighboring region in which the other sublattice is favored. By occupying different sublattices at different regions, the total energy of the resulting globally disordered ground states can be considerably lower than those of a perfect single-sublattice crystal (namely, energy density less than $-1/2$). This property is visualized by the example of Fig.~\ref{fig-GS-2d-square}, and it is also clearly indicated in Fig.~\ref{fig-E_mean_vs_rho_d2}. The different locally ordered domains are separated by boundaries composed of empty unfrozen sites, leading to intricate mosaic-like patterns (Fig.~\ref{fig-GS-2d-square-3}).

To obtain a deeper understanding on the configuration subspace formed by all the ground-state packing configurations, we distinguish the sites of the unfrozen subsystem into three types: type-$1$, type-$0$, and type-$*$ (see Sec.~\ref{subsec:3type}). Figure~\ref{fig-GS-space_2D_joint} demonstrates that the fraction of type-$*$ unfrozen sites is less than $0.25$ for the whole range of explored initial packing densities $\rho \in [0.08, 0.24]$; most of the unfrozen sites are of type-$1$ and type-$0$, with type-$1$ sites more and more abundant in comparison with type-$0$ sites as $\rho$ increases. These results mean that most of the unfrozen sites have extremely strong polarization in their packing states; a randomly chosen site of the unfrozen subsystem has a high probability ($> 40\%$) to be occupied in all the ground states, and it has a slightly lower probability ($> 35\%$) to be empty in all the ground states. 

We next investigate how the type-$1$ and type-$0$ sites are distributed in the unfrozen subsystem. For the two checkerboard sublattices of the periodic square lattice, we refer to the one with more type-$1$ unfrozen sites as sublattice A and the other one as sublattice B. Let us denote by $N_A^1$, $N_A^0$ and $N_A^*$ the total number of type-$1$, type-$0$ and type-$*$ vertices in sublattice A of the unfrozen subsystem; similarly we have $N_B^1$, $N_B^0$ and $N_B^*$ for sublattice B. Notice that the total number of unfrozen sites in sublattice A is simply $N_A^1+N_A^0+N_A^*$, and similarly for sublattice B. By our convention we have $N_A^1 \geq N_B^1$. Figure~\ref{fig-GS-space_2D_AB} demonstrates the fractions of type-$1$, type-$0$ and type-$*$ sites within the sublattices A and B for the systems of side length $L=1000$ (results are quantitatively similar for larger system sizes). When the initial packing density $\rho$ is less than $0.09$, almost all the unfrozen sites of sublattice A are of type-$1$ and those of sublattice B are of type-$0$, indicating that the ground states are long-range ordered crystalline states.  These results are consistent with the known fact that there exists a crystalline phase at exactly $\rho = 0$ and sufficiently low chemical potentials $\mu$.

When $\rho$ becomes larger, the type-$1$ sites become more and more equally distributed in both sublattices (Fig.~\ref{fig-GS-space_2D_AB}), indicating that sublattice A becomes less and less dominant in the densest packing configurations. The ground states at $\rho > 0.1$ could be regarded as glass states, in the sense that some sites of sublattice A have extremely strong preference to be occupied (type-$1$) and some other sites of the same sublattice A have extremely strong preference to be empty (type-$0$).

\subsection{Apparent phase transition}
\label{subsec:2dpt}

To quantify the degree of order in the ground-state configuration space and the symmetry-breaking between the two checkerboard sublattices, we define an order parameter $m_I$ as
\begin{equation}
m_I \, = \, \frac{ (N_A^1 - N_A^0) - (N_B^1 - N_B^0)}{ N_A^1 +N_A^0 + N_B^1 +N_B^0 } \; .
\label{eq:order_parameter_imbalance}
\end{equation}
If most of the sites in sublattice A of the unfrozen subsystem are type-$1$ sites ($N_A^1 - N_A^0$ big) and most of the sites in sublattice $B$ are type-$0$ sites ($N_B^0 - N_B^1$ big), then $m_I$ will be approaching unity, indicating an globally ordered phase. On the other hand, if $(N_A^1-N_A^0)$ are comparable to $(N_B^1-N_B^0)$, then $m_I$ will be close to zero, indicating no dominance of one sublattice. We refer to $m_I$ as the imbalance order parameter. It is similar to the conventional magnetization order parameter $m$ of the Ising model. 

\begin{figure}
\centering
\includegraphics[width=0.6\linewidth]{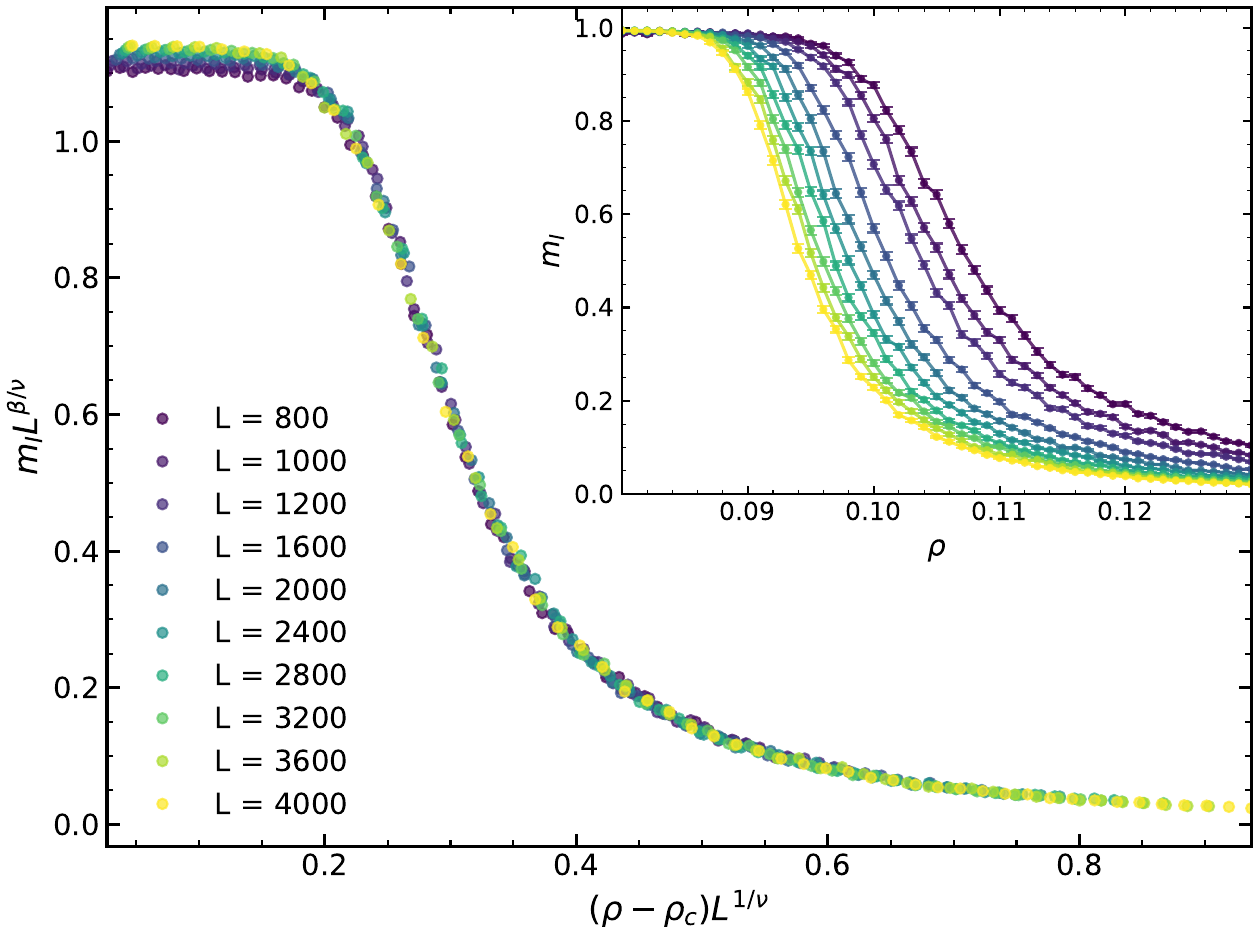}
\caption{
Imbalance order parameter $m_I$ versus the fraction $\rho$ of initially occupied sites with periodic square lattice with side length from $L=800$ to $L=4000$. The main figure shows the 
finite-size scaling relationship between the rescaled imbalance order parameter $m_IL^{\beta/\nu}$ and the rescaled initial packing density $(\rho-\rho^*)L^{1/\nu}$, with $\rho^*=0.0774(4)$, $\nu= 2.88(3)$ and $\beta=0.05(1)$. The chi-square value of the fitting is $\chi^2=1.39$ with 260 degrees of freedom for the polynomial fit. The raw data of $m_I$ are shown in the inset. Each data point is the average over $1000$ independent samples of the initial configuration. 
}
\label{fig:mI2d}
\end{figure}

The inset of Fig.~\ref{fig:mI2d} draws the $m_I$-versus-$\rho$ curves for various different system sizes ranging from $L=800$ to $L=4000$. These results suggest that the ground states are ordered at $\rho < 0.08$ for these finite systems.  We also notice that as $L$ increases the function $m_I(\rho)$ gradually shifts to the left, indicating the parameter space of the ordered phase shrinks as the system increases.

To investigate the possibility of a phase transition, pinpoint its location should it exist and determine the associated critical exponents, we conduct a finite-size scaling analysis of the ground-state $m_I$ results obtained on the square lattices. The critical behavior of a system near a continuous phase transition is governed by a power law relationship characterized by critical exponents. For a system of finite size $L$, the imbalance order parameter $m_I$ is expected to follow the scaling form~\cite{Binder-1981}
\begin{equation}
 m_I  \,= \,  L^{-\beta/\nu} f\bigl( (\rho - \rho^*) L^{1/\nu} \bigr) \; .
\label{eq:fss_ansatz_imbalance}
\end{equation}
In our FSS analysis, the fitting function $f(x)$ is approximated by a high-order polynomial (\ref{eq:poly}) with an optimal value of $q \approx 12$, and we minimize the chi-square deviation (\ref{eq:chisq_red}) between the numerical simulation data and this scaling relationship. 

Our FSS numerical results are shown in Fig.~\ref{fig:mI2d}, with critical density $\rho^* =0.0744$ and two critical exponents $\nu=2.88$, $\beta = 0.05$. Notice that this predicted critical density value $\rho^*$ is located below all the density values of the samples explored in our numerical experiments ($\rho \geq 0.08$), so this value can only be regarded as an extrapolation. Notice also that the chi-square value of the fitting $\chi^2 = 1.39$ is considerably higher than unity, suggesting that the scaling form (\ref{eq:fss_ansatz_imbalance}) is not yet strongly supported by the simulation data. The relatively poor fitting is also manifested in Fig.~\ref{fig:mI2d}, as the different curves of $m_I L^{\beta/\nu}$ versus $(\rho-\rho^*) L^{1/\nu}$ fail to superimpose onto each when the packing density is approaching $\rho^*$ from above. There may be extremely strong finite-size effect in the two-dimensional systems, and the estimated critical point $\rho^* \approx 0.07$ between the crystalline and glass phases of the ground-state configuration space may only valid for relatively small system sizes $L$. We expect that when the system sizes $L$  further increase, the predicted critical densities $\rho^*$ will shift to even smaller values.

The data collapse of Fig.~\ref{fig:mI2d} therefore only offers us an upper bound on the true critical packing density $\rho^*$.  In Sect.~\ref{sec:ImariMa} we will argue that the true phase transition point for infinite-size systems should be located at exactly $\rho^* = 0$.

\section{Frozen sites destroy long-range order in the square lattice}
\label{sec:ImariMa}

For the square lattice systems, our numerical results of Sect.~\ref{section:phasetransition2D} did not rule out the possibility that the ground states of the unfrozen subsystem are ordered crystalline states at small positive values of initial packing densities $\rho$. Would there be a true phase transition at a positive value of $\rho$ located in the range of $\rho \in (0, 0.08)$, when the system size $L$ goes to infinity? We now present further numerical evidence to support the conjecture of disordered ground states for $D=2$ systems at any positive value of $\rho$. 

\begin{figure}
\centering
\includegraphics[width=0.6\linewidth]{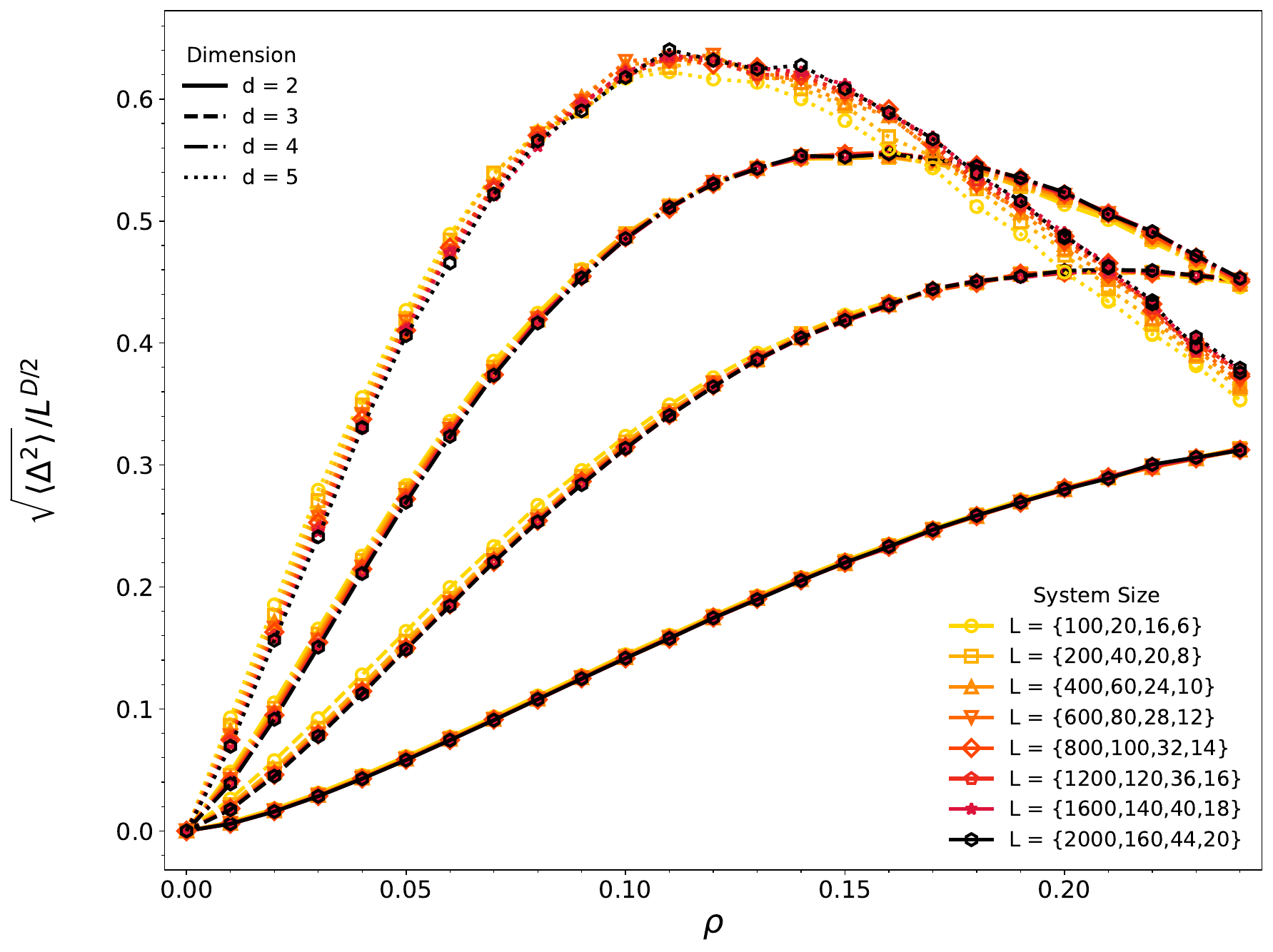}
\caption{
Standard deviation of the number imbalance $\Delta$ of hypercubic region of side length $L$, scaled by $L^{D/2}$, versus the density $\rho$ of initially occupied sites. Data are shown for dimensions $D=2, 3, 4, 5$. The different symbols and colors correspond to sets of side lengths $L$, where the values in each list correspond to dimensions in ascending order. Each point is calculated from $10^5$ independent samples of initial packing configurations. 
}
\label{fig:delta_e_scaling}
\end{figure}

Following the well-known theoretical argument of Imry and Ma on the issue of long-range order in the random-field Ising model~\cite{Imry-Ma-1975}, we analyze the energy change associated with flipping a whole local region of the unfrozen subsystem. In our numerical experiments, performed on the $D$-dimensional hypercubic lattices with $D\in \{2,3,4,5\}$, we focus on a fixed hypercubic region of side length $L$ within a much larger hypercubic lattice with periodic boundary condition. The total number of sites in this focused region is $L^D$. We then sample a large number $M = 10^5$ random initial packing configurations for the whole system at fixed initial packing density $\rho$. For each of these initial configurations, we get the corresponding unfrozen subsystem, and then count how many sites of the checkerboard sublattices $A$ and $B$ of the focus region belong to the giant connected component of this unfrozen subsystem. We denote these two numbers of unfrozen sites as $N_A$ and $N_B$, respectively. By this way we get $M$ sampled data pairs $(N_A, N_B)$ for the focus region. Let us define the number imbalance as
\begin{equation}
\Delta \, := \,  N_A - N_B \; .
\label{eq:numdiff}
\end{equation}
The expected value of $\Delta$ is obviously zero. The standard deviation of $\Delta$ is easily estimated from these samples, and the results are shown in Fig.~\ref{fig:delta_e_scaling}. For all the checked dimensions $D$ we find that this standard deviation is proportional to the square root of the number of sizes $L^{D/2}$ of the focus region, namely
\begin{equation}
\bigl\langle  \Delta^2  \bigr\rangle \, = \, \sigma_0(\rho) L^D \; ,
\label{eq:numdiffsquare}
\end{equation}
where $\sigma_0(\rho)$ is $\rho$-dependent numerical coefficient. 

With this empirical observation (\ref{eq:numdiffsquare}), we can now answer the following question: Suppose the giant connected component of the unfrozen subsystem of an infinite hypercubic lattice is taking a long-range ordered packing configuration (say, sites of sublattice $A$ are all being occupied and sites of sublattice $B$ are all empty), could this crystalline configuration be possibly stable against flipping the states of all the unfrozen sites of a hypercubic region of large side length $L$? Such a flipping will cause an energy increase $\Delta E \approx 2 D L^{D-1}$ (when $\rho$ is very close to zero) at the boundary of this flipped region, but it may cause an energy decrease  $\Delta$ (if $\Delta = N_A - N_B$ is negative) in the internal of the region. Assuming $\Delta$ obeys the Gaussian distribution with mean zero and variance $\sigma_0 L^D$, the probability $w$ of a net decrease in energy is then
\begin{equation}
\begin{aligned}
 w \, & = \, \int_{-\infty}^{- 2 D L^{D-1}} \textrm{d} \Delta \frac{1}{\sqrt{2 \pi \sigma_0 L^D}} \exp\Bigl( - \frac{\Delta^2}{2 \sigma_0 L^D} \Bigr) \\
 & = \, \frac{1}{2} \textrm{erfc}\Bigl(  \sqrt{\frac{2 D L^{D-2}}{\sigma_0}} \, \Bigr) \; ,
 \end{aligned}
 \label{eq:winprob}
\end{equation}
where $\textrm{erfc}(z)$ is the complementary error function, which has the asymptotic form of $\textrm{erfc}(z) \approx e^{-z^2}/(\sqrt{\pi} z)$ when $z$ is much larger than unity.

If $D=2$, then we obtain that the probability $w = \frac{1}{2}\textrm{erf}(\sqrt{4/\sigma_0})$ is of order unity for all region sizes $L$. This means that within an infinite square lattice there exist extremely many square regions of even very large side lengths $L$, the flipping of which could increase the number of occupied sites (lower the energy). Then a ground state of an infinite square lattice can not be long-range ordered for any positive value of $\rho$. In other words, the ground states of the unfrozen subsystem must be globally disordered at any $\rho > 0$ in the limit of infinite systems~\cite{Imry-Ma-1975,Aizenman-Wehr-1990,Ding-Zhuang-2024,Bricmont-Kupiainen-1988}.

\begin{figure*}
\centering
\subfigure[]{
\includegraphics[width=0.48\textwidth]{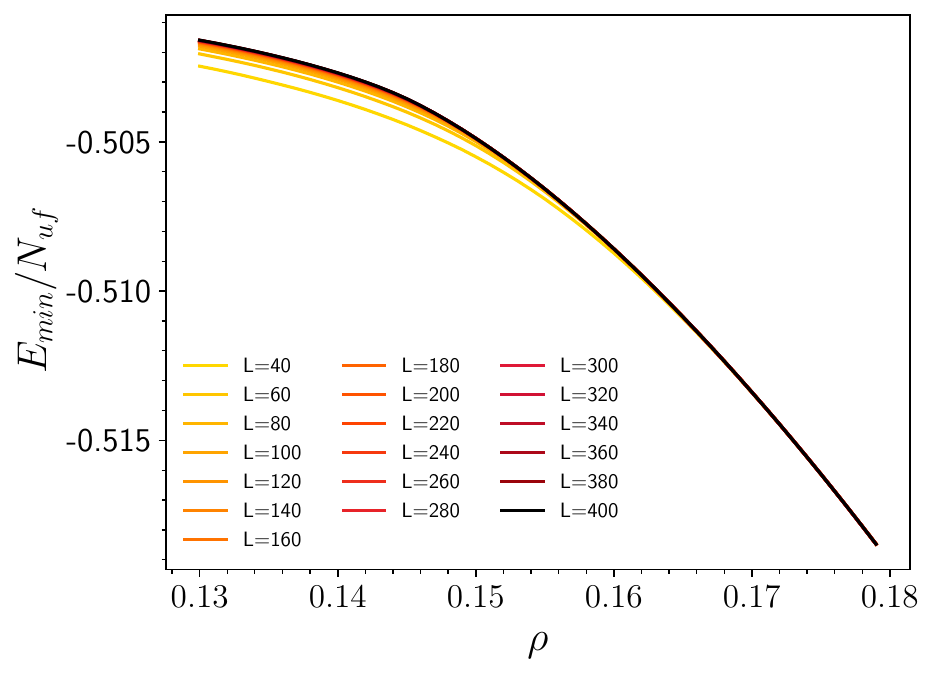}
\label{fig-E_mean_vs_rho_d3}
}
\subfigure[]{
\includegraphics[width=0.46\textwidth]{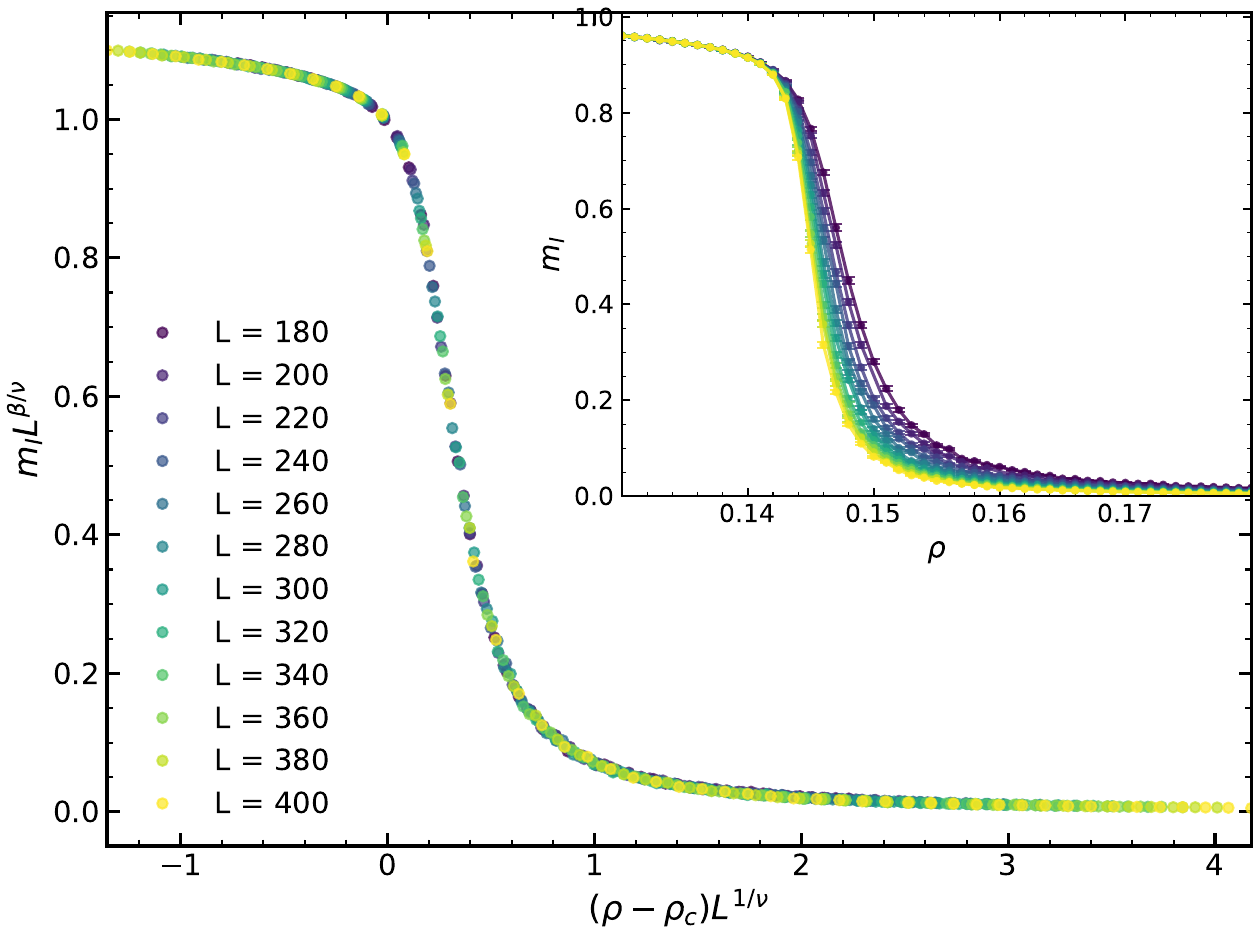}
\label{fig:fss_collapse_imbalance3d}
}
\caption{
Numerical results for periodic cubic lattices of various side lengths $L$. 
(a) The ground-state energy density $E_{\textrm{min}}/N_{\textrm{uf}}$ of the unfrozen subsystem versus the initial packing density $\rho$. Each data point is the average over $1000$ independent random initial packing configuration; shaded regions represent standard errors of the mean. (b) Finite-size scaling analysis on the imbalance order parameter $m_I$ (raw data shown in inset). Data from different system sizes $L$ fall onto a single curve of $m_I L^{\beta/\nu}$ versus $(\rho - \rho^*) L^{1/\nu}$, with $\rho^* = 0.14226(4)$, $\nu = 1.27(2)$ and $\beta = 0.0288(4)$. The chi-square value of the fitting is $\chi^2 = 1.0127$ with 130 degrees of freedom. 
}
\end{figure*}

For $D \geq 3$ the situation is qualitatively different. We see that $w$ approaches zero as $L$ increases. Therefore each flipped region must be of small size and the global long-range order and the dominance of one checkerboard sublattice will be persistent as long as $\rho$ is not too large (at $\rho=0$ the two ground states of the system are surely long-range ordered). In agreement with this local stability argument, the results of the next section indeed confirm that ordered ground states will persist to small positive values of $\rho$ for higher-dimensional systems. For the infinite $D=3$ cubic lattice, long-range order of the ground states can extend up to $\rho^* \approx 0.142$.

\section{Ground states of cubic lattices}
\label{section:phasetransition3D}

Following the same procedure of Sect.~\ref{section:phasetransition2D}, we study the ground-state properties of the periodic cubic lattices ($D=3$). Each site in the lattice has six nearest neighbors. The mean ground-state energy density of the unfrozen subsystem, $E_{\textrm{min}}/N_{\textrm{uf}}$, as a function of initial packing density $\rho$ is shown in Fig.~\ref{fig-E_mean_vs_rho_d3} for lattice side lengths ranging from $L=40$ to $L=400$. Starting from the maximum value of $-1/2$ at $\rho = 0$,  the mean ground-state energy density decreases with $\rho$ in a continuous and smooth manner within the percolating phase of the unfrozen subsystem ($\rho <  0.28$). This is qualitatively identical to Fig.~\ref{fig-E_mean_vs_rho_d2} of the $D=2$ systems.

\begin{figure*}
\centering
\subfigure[]{
\includegraphics[width=0.47\textwidth]{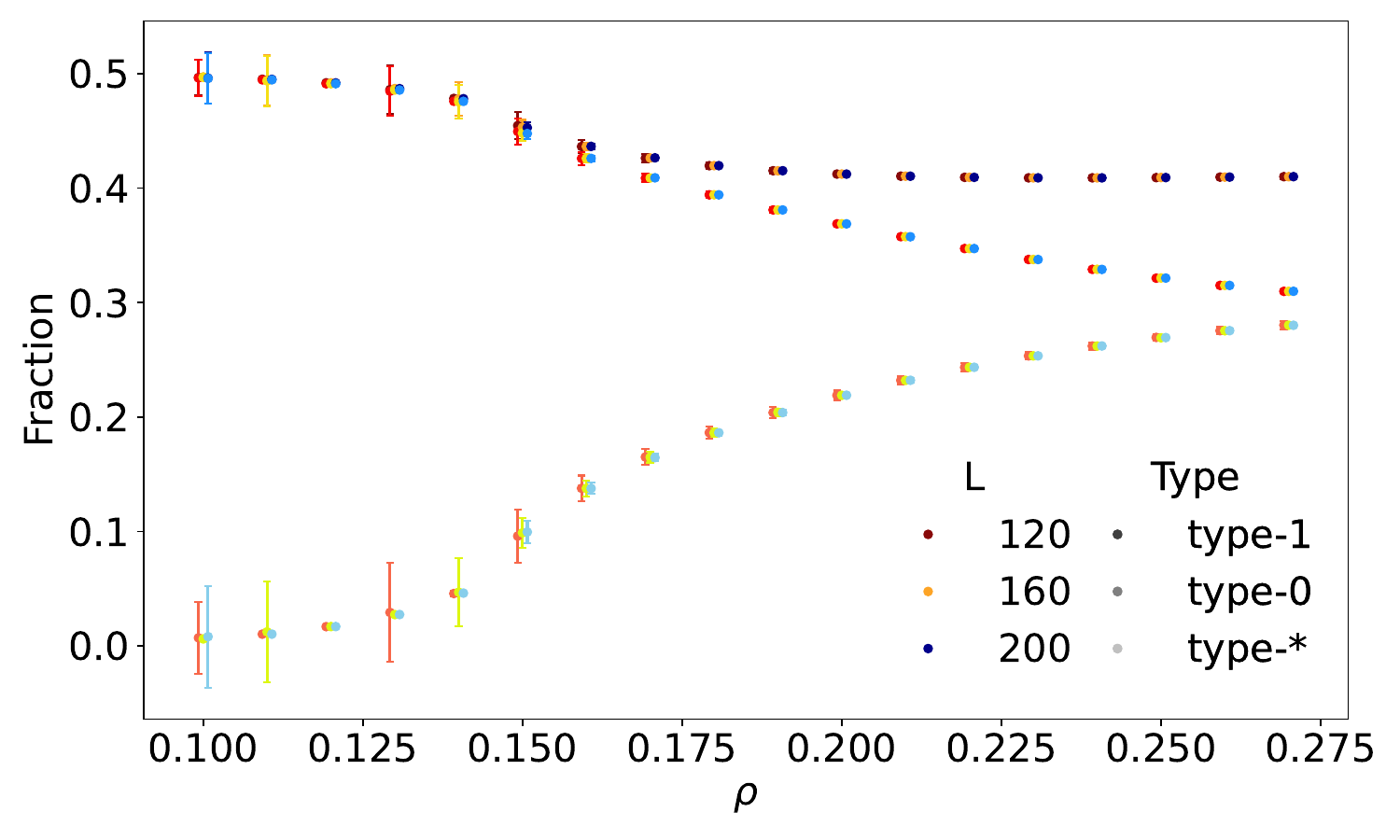}
\label{fig-GS-space_3D_joint}
}
\subfigure[]{
\includegraphics[width=0.47\textwidth]{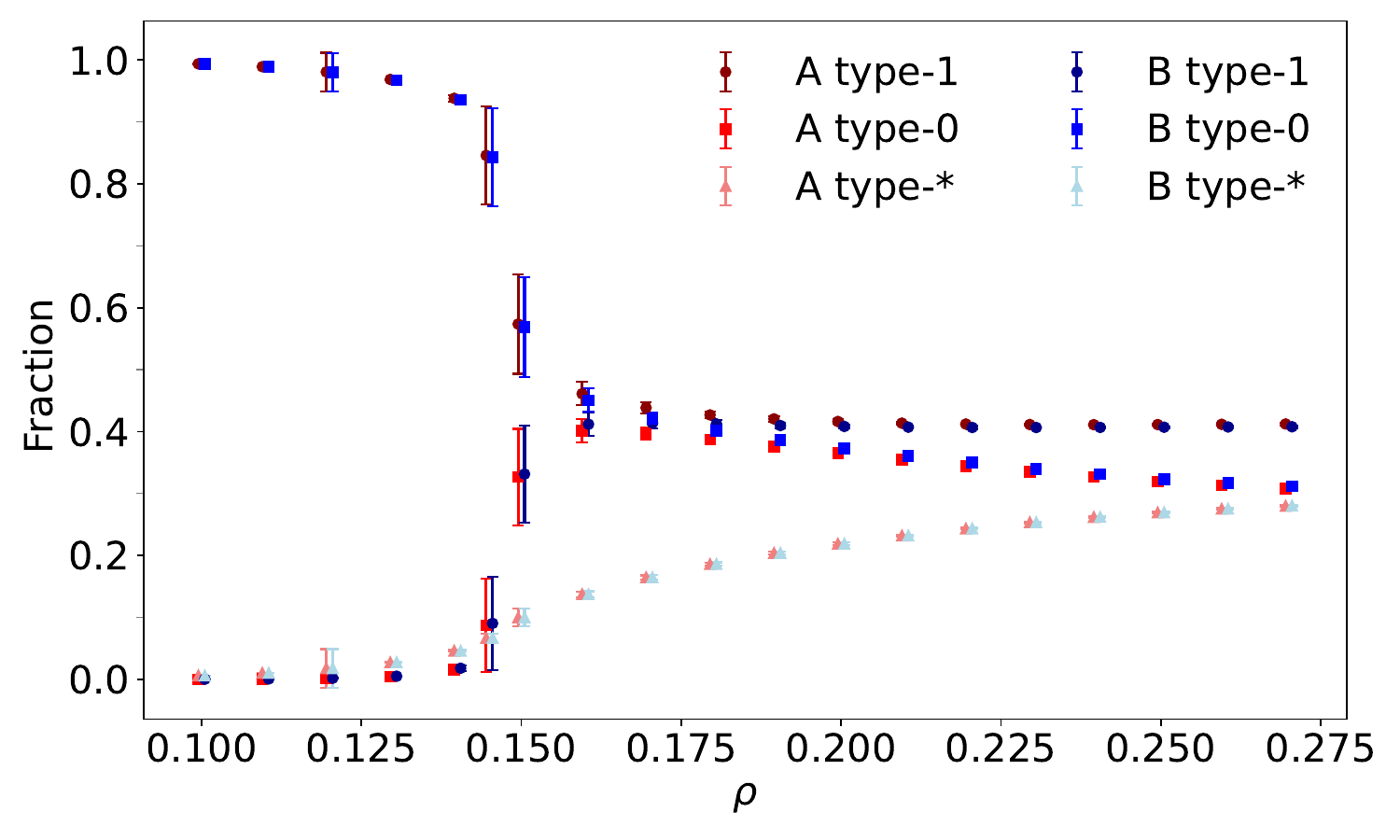}
\label{fig-GS-space_3D_AB}
}
\caption{
Fractions of sites with the three different coarse-grained states (type-$1$, type-$0$, and type-$\ast$) in the giant connected component of the  unfrozen subsystem on periodic cubic lattice. (a) Results for different system sizes $L= 120, 160, 200.$ Different system sizes are shown in different colors, and within each size, different type fractions are shown in varying shades of the same color. (b) Results with both sublattice A and sublattice B for $L = 200$. Small horizontal offset ($0.0005$) is applied to the data points at the same $\rho$ for clarity.
}
\label{fig-GS-space_3D}
\end{figure*}

The fractions of the type-$1$, type-$0$ and type-$*$ unfrozen sites in the giant connected component of the unfrozen subsystem are shown in Fig.~\ref{fig-GS-space_3D}, which exhibits some similarity with Fig.~\ref{fig-GS_space_2D} of the square lattices. From  we see that, As the initial packing density $\rho$ increases from $0.14$ to $0.16$, the fraction of type-$1$ sites drops from a very high level of $\approx 0.5$ to a lower level of $\approx 0.4$; while the fraction of type-$0$ sites shows a more gradual decreasing trend in $\rho \in [0.14, 0.275]$, and this decreasing trend is compensated with the continuous increasing of the fraction of type-$*$ sites (Fig.~\ref{fig-GS-space_3D_joint}). When $\rho \leq 0.14$, almost all the unfrozen sites of sublattice A are type-$1$ occupied sites while almost all the unfrozen sites of sublattice B are type-$0$ empty sites (Fig.~\ref{fig-GS-space_3D_AB}), suggesting the strong dominance of sublattice A and the crystalline order of the whole system. However, there appears to be a sudden drop in the dominance of sublattice A at the vicinity of $\rho = 0.15$, indicating the existence of a phase transition (Fig.~\ref{fig-GS-space_3D_AB}).

For $\rho > 0.15$ the fractions of type-$1$ sites of the two sublattices are equal to each other; and this property also holds for the fractions of type-$0$ and type-$*$ sites. Qualitatively similar to the disordered two-dimensional ground-state configurations (Fig.~\ref{fig-GS-2d-square}), the type-$1$ sites of  both sublattices of the three-dimensional systems form many local domains, and the interfaces between type-$1$ domain of sublatice A and the adjacent type-$1$ domains of sublattice B are highly irregular in shape.

We estimate the imbalance order parameter $m_I$ as defined in Eq.~(\ref{eq:order_parameter_imbalance}) for various initial densities and system sizes (Fig.~\ref{fig:fss_collapse_imbalance3d}, inset). By performing a finite-size scaling analysis under the scaling ansatz (\ref{eq:fss_ansatz_imbalance}), we find that all the raw numerical data can be perfectly collapsed onto a single curve (Fig.~\ref{fig:fss_collapse_imbalance3d}). We obtain $\rho^* = 0.14226$ for the critical packing density, and scaling exponents $\nu = 1.27$, $\beta = 0.0288$. The chi-square value $\chi^2 = 1.0127$ of this fitting is very close to unity, indicating that the scaling property (\ref{eq:fss_ansatz_imbalance}) is valid for the cubic lattice systems at the vicinity of $\rho = \rho^*$. Notice that the predicted critical point $\rho^*$ is located well within the range $\rho \in [0.135, 0.160]$ of raw numerical data, so it is no longer an extrapolation as in the $D=2$ fitting of Fig.~\ref{fig:mI2d}.

Our estimated value for the critical exponent $\nu$ is close to the results of $\nu \approx 1.38(10)$ in the recent literature on the random-field Ising model (RFIM) and the estimated extremely small value of $\beta$ is also consistent with the RFIM result of $\beta \approx 0.056$~\cite{Hartmann-Young-2001,Middleton-2001a,Dukovski-Machta-2003,Wu-Machta-2006,Fytas-MartinMayor-2016b}. Indeed the kinetic systems studied here are physically similar to the RFIM. To avoid a domain boundary, two neighboring regions of the giant connected component of the unfrozen subsystem are encouraged to choose the same packing preference (sublattice $A$ or sublattice $B$). This alignment can be mapped to the ferromagnetic interaction between two neighboring spins of the RFIM. The individual kinetically frozen domains impose local biases (some toward sublattice $A$ and the others toward sublattice $B$) to the local packing patterns, which may be mapped to the random local fields of the RFIM. We can take advantage of this similarity to understand more deeply the theoretical aspect of the ground-state phase transition~\cite{Chen-etal-2025}.

It remains to be checked whether the present kinetic spin systems belong to the same universality class of the RFIM. Given the correlated nature of the frozen sites in the kinetic systems, we would not be surprised if they belong to a slightly different universality class.  The extremely small value of critical exponent $\beta\approx 0.03$ also means that more work needs to be done to fully understand the nature of the ground-state  transition at $\rho \approx 0.142$ between the ordered crystalline phase and the disordered glass phase (see, e.g., Ref.~\cite{Liu-etal-2025} for a closely relevant recent work). We have performed finite-size scaling fitting by fixing $\beta = 0$, but the result was much worse; and we have also checked that the distribution of the sampled imbalance order parameters $m_I$ has only a single peak. It may be safe to interpret the small positive value of $\beta$ as a signiture of a highly cooperative but still continuous crystal-to-glass phase transition of the ground-state configurations.

\section{Conclusion and discussions}
\label{section:conclusion}

In summary, we studied some of the thermodynamic properties of the most severely constrained ($K=1$) Fredrickson-Andersen kinetic spin model on finite-dimensional lattices by computer simulations. Starting from a random initial configuration with packing density $\rho$, the FA kinetic rule renders some of the lattice sites being permanently frozen in their initial packing states, and it also leads to effective excluded-volume (hard-core) interactions between nearest-neighboring unfrozen sites. We demonstrated that the subsystem formed by all the unfrozen sites and the bonds between them experiences a continuous collapse phase transition as the initial packing density $\rho$ increases from zero to certain critical value $\rho_c$. This collapse transition belongs to the universality class of the conventional site percolation transition. The giant connected component of the unfrozen subsystem ceases to be extensive in size as $\rho$ exceeds $\rho_c$, with $\rho_c = 0.2475$ for the two-dimensional square lattice and $\rho_c = 0.2809$ for the three-dimensional cubic lattice.

We then investigated the densest packing configurations (the ground states) of the unfrozen subsystem in the percolating phase with $\rho$ much below $\rho_c$.  For the three-dimensional cubic lattice, we demonstrated that there is a  phase transition in the ground states at the critical density $\rho^* = 0.1423$; the ground states are long-range ordered (the crystalline phase) if $\rho < \rho^*$ and they are locally ordered by  long-range disordered (the glass phase) if $\rho > \rho^*$; the phase transition may likely be continuous in nature but the estimated critical exponent $\beta \approx 0.03$ is very small. For the two-dimensional square lattice, our numerical results are consistent with the prediction of $\rho^* = 0$, namely that the ground states are long-range disordered for any positive $\rho$ and they are ordered only at $\rho = 0$. (By improving our theoretical arguments presented in Ref.~\cite{Chen-etal-2025}, we are able to rigorously prove that $\rho^* = 0$. This mathematical proof will be described in a separate paper.)

Taken together, our work  clearly confirmed that the very severe kinetic constraint with $K=1$ can induce thermodynamic phase transitions in finite-dimensional kinetic models. The underlying physics may be very simple:  Kinetically frozen sites serve as randomly located  defects of the lattice,  affecting it structural connectivity and inducing strong local biases on the dense packing patterns of the two competing sublattices.

The ground-state phase transition between the crystalline phase and the glass phase corresponds to chemical potential $\mu = -\infty$. We expect that the low-defect ($\rho < \rho^*$) crystalline phase will be stable up to certain $\rho$-dependent critical chemical potential $\mu_1(\rho)$. The critical chemical potential $\mu_1(0) \approx -2.406$ at $\rho=0$ has already been exactly known for the $D=2$ hexagonal lattices~\cite{Baxter-1980}, and for the $D=3$ cubic lattice the critical chemical potential has been numerically determined  as $\mu_1(0)\approx -0.0552$~\cite{CunhaNetto-Dickman-2011,Panagiotopoulos-2005,Heringa-Blote-1996}. The whole critical line $\mu_1(\rho)$ awaits to be worked out by numerical simulations.  The high-defect ($\rho > \rho^*$) glass phase may also persist up to certain $\rho$-dependent critical chemical potential $\mu_2(\rho)$, which marks a phase transition to the disordered gas phase. There may exist a tricritical point at the intersection of the crystalline, glass, and disordered gas phases, similar to the case of the $\pm J$ random-bond Ising models with adjustable fraction of ferromagnetic interactions~\cite{Liu-etal-2025}. Extending the computation to the construction of the whole phase diagram of the $K=1$ kinetic spin model is a time-consuming task for future investigations.

Our numerical results suggested that the critical initial occupation density $\rho^*$ of the thermodynamic phase transition in the ground states is strictly positive for the $K=1$ three-dimensional FA kinetic models, but this critical density is exactly zero for the two-dimensional systems. Further theoretical efforts are needed to characterize the universality classes of these continuous phase transitions and to understand the dimension effect. The frozen sites induced by the local kinetic rule serve as  defects in the lattice, and they impose local biases to the adjacent unfrozen sites of the lattice. Theoretical insights and methods developed from studying the random field Ising model can be adapted to studying this kinetic model system~\cite{Imry-Ma-1975,Aizenman-Wehr-1990,Ding-Zhuang-2024,Bricmont-Kupiainen-1988}. We are making progress along this direction~\cite{Chen-etal-2025}. We also notice that, in recent years the $M$-layer construction has been introduced to study finite-dimensional corrections to the Bethe lattice results and to build $\epsilon$-expansion for the critical exponents~\cite{Rizzo-2018,Angelini-Palazzi-etal-2025,Angelini-Palazzi-etal-2025b}.

We have distinguished the unfrozen lattice sites into three types, with  type-$1$ and  type-$0$ sites having fixed states among all the ground states of the unfrozen subsystem. When the chemical potential $\mu$ is very negative but finite, type-$1$ and type-$0$ sites will also be able to change states, but their  dynamical properties may be very different from those of the type-$*$ sites.  This possible issue of dynamical heterogeneity is worthy to be thoroughly explored in the near future. As type-$1$ and type-$0$ unfrozen sties become thermally exciable at finite $\mu$ values, whether the crystal-to-glass phase transition still persists or changes to be a crossover is also an open issue to be studied.

We have completely neglected dynamical issues (such as facilitation, growing relaxation times, and dynamical heterogeity, which are widely emphasized in the structural glass literature) in the present work. It may be very interesting to explore the dynamical consqeuences of the reported static, ground-state structural transition, especially at the vicinity of $\rho^*$ at finite chemical potentials.

Our present work only addressed the extreme case of $K=1$, for which the kinetic constraint induces local excluded-volume effect in the lattice. The local constraint of the $K=1$ systems can be regarded as a limiting case of the topological constraint of $K$-core absence~\cite{Zhou-2024}. New thermodynamic features may emerge if we relax the kinetic constraint to $K=2$. For this latter case, the frozen occupied sites organize into individual closed loops and they form a $2$-core of initially occupied sites. These frozen sites may also induce a collapse transition in these $K=2$ systems. The occupied sites of the microscopic configurations of the corresponding unfrozen subsystem are subject to a global topological constraint of $2$-core absence and therefore they must organize into tree structures free of any closed loops~\cite{Zhou-2024}. A ground state of the unfrozen subsystem is then a densest packing pattern of mutually repelling tree structures of various sizes and shapes~\cite{Zhou-2013,Zhou-2022}. These tree structures have much higher internal entropy and they may greatly enhance the stability of the disordered packing configurations. These finite-dimensional $K=2$ kinetic lattice systems present greater challenges for computer simulations and theoretical analysis but they will shed deeper insights into the nature of structural glasses. We are working on an efficient configuration-sampling algorithm for these $K=2$ systems. We have accumulated extensive numerical results for the simplest initial condition of $\rho=0$ (all the sites being unfrozen) and will report these results in an accompanying paper.

\section*{Data availability}

The raw data and the codes accompanying this manuscript can be accessed at here~\cite{github}.

\begin{acknowledgments}
We thank Professor Wei Wei for very helpful discussions. The following funding supports are acknowledged: National Natural Science Foundation of China Grants No.~12247104, No.~T2541021  and No.~12447101. Numerical simulations were carried out at the HPC cluster of ITP-CAS and also at the BSCC-A3 platform of the National Supercomputer Center in Beijing.
\end{acknowledgments}

\begin{appendix}

\section{Cycle simplication on the state-flexible subgraph}
\label{app:cs}

An important concept,  cycle simplification, was introduced by Wei and co-authors to treat the state-flexible subgraph of a bipartite graph~\cite{Wei-etal-2015,Wei-Zhang-etal-2015}. A final reduced  bipartite subgraph will be reached by applying the cycle simplification process repeatedly, starting from the initial state-flexible subgraph. This cycle simplication process is very useful for the purpose of sampling densest packing configurations uniformly at random for the original state-flexible subgraph.

The cycle simplification process goes as follows: If there is a closed cycle of alternating matching  and unmatched edges, e.g., the cycle formed by the set of edges
$$\{(e,x), (x, f), (f, w), (w, g), (g, v), (e, v)\}$$
of Fig.~\ref{f120226}, then all the sites  of sublattice $A$ in this cycle will take the same state and all the sites of sublattice $B$ in this cycle will take the same opposite state in any ground state, so we can replace the whole cycle by a single matching edge of this cycle (e.g., edge $(e, x)$) and delete all the other edges of this cycle. If there is an edge (e.g., edge $(h, v)$ of Fig.~\ref{f120226}) linking between this deleted cycle and a site not belonging to this cycle, then the end point of this edge in this cycle is replaced by the equivalent site ($x$) of the representing edge. This cycle simplification process is then applied again on the simplified bipartite subgraph if it still contains at least one cycle of alternating matching and unmatched edges. The final reduced bipartite subgraph then has no any cycle of alternating matched and unmatched edges. Notice that it may still contains cycles, but such cycles are not formed by matching and unmatched edges in alternation.

It was proven in Ref.~\cite{Wei-etal-2015} that, after cycle simplification, the resulting reduced bipartite subgraph of the original state-flexible subgraph contains no leaf-removal core (all the sites  can be removed through the leaf-removal process). This is easy to prove. If there is a matched edge $(i_a, j_a)$ in this graph with  $i_a$ being a leaf site ($i_a$ has only one attached edge which is $(i_a, j_a)$), then we can apply the leaf-removal process on the edge $(i_a, j_a)$ and delete it together with the two end sites $i_a$ and $j_a$ from the subgraph.  We keep performing this leaf-removal process as long as at least one of the remaining matched edges is a leaf edge. If there is a non-empty final bipartite subgraph (a leaf-removal core) surviving this leaf-removal iteration, then both end sites  of any surviving matching edge must be attached by at least one unmatched edge, and a cycle of alternating matching and unmatched edges can then be constructed in the surviving subgraph. But such a cycle has already been deleted during the cycle simplification process and it can not exist in the final reduced subgraph. Therefore the final reduced subgraph must not contain a leaf-removal core.

The absence of leaf-removal core means that there is no long-range frustration in the final reduced state-flexible bipartite subgraph~\cite{Zhou-2005a}. The configuration space formed by all the densest packing configurations is ergodic even if only single-site state flip operations are employed to walk through this space~\cite{Wei-Zhang-etal-2015}. It is easy to see that there is one-to-one correspondence between a densest packing configuration of the reduced state-flexible subgraph and the original state-flexible subgraph.

\end{appendix}


%

\end{document}